# Light-Matter Interactions in Photosynthetic Protein Attached to solids and Nanostructures.

Itai Carmeli[1], * and Chanoch Carmeli[2]

[1]Faculty of Engineering, Bar-Ilan University, Ramat-Gan, 52900, Israel; [2]Department of Biochemistry and Molecular Biology, Tel Aviv University, Tel Aviv 69978 Israel.

*Corresponding author: icarmeli18@gmail.com

**Running Title**: Light-matter interactions in photosynthetic proteins

**ABSTRACT**
The interaction of light with photosynthetic proteins is an extremely efficient process and has been thoroughly investigated. However, exploring light-matter interactions in hybrid nano-solid-photosynthetic proteins is a relatively new and existing field of research. The properties of these hybrid materials significantly influence the energy levels, non-radiative energy transfer, absorption, and fluorescence of the photosynthetic proteins upon interaction with light. There is special interest in levering these light-matter interactions for applications such as photo-sensing and converting light energy to electricity. The development of efficient devices requires the formation of a junction for oriented attachment, facilitating efficient energy and electronic transfer between the solids and the proteins. This review will outline the major advancements in solid-state photosynthetic protein devices, elucidate the underlying mechanism, and assess electron transfer efficiency. Furthermore, it will explore and analyze the effect of plasmons on the enhancement of absorption, fluorescence, and photocurrent in hybrid devices.

**INTRODUCTION**
In the introduction, we will explore the background surrounding the constituent elements contributing to establishing efficient light-matter interactions and evaluating energy and electron transfer efficiency within the hybrid system. These elements encompass 1. the diverse array of photosynthetic proteins utilized in optoelectrical devices, 2. the various types of nanoparticles integrated into the hybrid system, 3. the mechanism of interactions between proteins and solid materials, 4. the characteristics defining photosynthetic protein-solid hybrids, 5. the different categories of electrochemical devices, and 6. the diverse array of solid-state devices involved.

**Photosynthetic proteins that were applied in optoelectrical devices**
Light-matter interactions profoundly affect the solid protein hybrids' energy levels, non-radiative energy transfer, absorption, and fluorescence properties [1, 2]. Of special interest is the light-matter interaction of the efficient photoactive photosynthetic proteins that can be used in photo-sensing applications and potentially in light energy conversion to electricity. The better studied among these photosynthetic proteins are reaction centers from plants, algae, and cyanobacteria: photosystem I (PSI) and photosystem II (PSII), the photosynthetic bacterial reaction centers (RC), chloroplasts light-harvesting chlorophyll protein complexes I (LHCI) and II (LHCII) and bacterial light-harvesting complexes (LH1-4). To fabricate devices, the proteins have to be immobilized on solid surfaces. The challenge here is to form ordered and oriented molecular films with good electronic coupling to the solid-state surface.

  1. **Nanoparticles used in hybrid systems**
Nanoparticles' shape and size enable the control of their optical spectrum and transient response. The coupling of metamaterials and plasmonic nanostructures to these proteins allowed tuning their electronic band structures and transient response. In turn, the bound proteins tune the plasmonic and optical properties of nanoparticles, nano-slits, and other exotic geometrical arrays. Non-radiative energy transfer: "can be enhanced through the hybridization of electronic transitions". The hybridization generates new eigenstates and can be delocalized over many molecules [1]. Importantly, the hybridization generates new light-matter states because of changes in the zero-point energies even without light [3].

  2. **Interaction between proteins and solids**
The study of the interaction between biological proteins and solid-state substrates, in general, has gained wider interest in recent years. It is reasonable to assume that proteins, that were perfected over hundreds of millions of years of evolution to become highly efficient catalysts, developed facile electron tunneling and highly efficient light energy conversion can contribute to new and superior properties of hybrid systems when compared to the properties of its components. Protein- solid hybrids were used in the development of devices in various scientific fields such as organic electronics [4], bioelectronics [5], biosensors [6], and light energy conversion to electricity [7-20]. The various reviews analyze the methods for the fabrication of photovoltaic devices based on photosynthetic protein and the ways to enhance their efficiency and energy output. In this review, we aim to analyze light-matter interactions between the solid-state material and the photosynthetic proteins and the role of electronic junctions in the efficiency of the devices. We will also point out the advantages and shortcomings of solid-state and electrochemical devices. Electrochemical devices containing intact bacteria and algae were not covered in this review

### 3. Photosynthetic protein-solid hybrids

The study of photosynthetic proteins in solid-state hybrids was motivated by the desire to apply their superior optoelectronic properties in the fabrication of solid devices. The results of these studies indicated that these proteins have the potential to enhance antenna properties, serve as light and herbicide sensors, be applied to photovoltaics devices, used as photogates in field effect transistors, serve in light modulation, and be used as spin injectors. Types of solid-state materials utilized in solid-state-protein hybrids included metal, semiconductors, carbon nanotubes, graphene, metamaterials, nanoparticles, and plasmonic microstructures. The devices were constructed using meso, micro, and nano-size technologies. In the search for the interactive interface between the proteins and the solids, a variety of technologies were applied. Most prominent among them are electrostatic and hydrophobic interaction, phi-phi interactions, covalent binding, tag-metal ion junctions, and molecular wire substitutions. The advantages and shortcomings of the various approaches for the formation of an effective interface will be discussed.

### 4. Electrochemical devices

Electrochemical techniques were applied in many cases; however, more recently solid-state devices were developed. The advantage of electrochemical devices is the presence of an aqueous environment favorable to protein function, the possible use of partially oriented proteins, and the use of mobile electron carriers to mediate current between the proteins and the electrodes. Yet, the proteins are relatively unstable because of degradation by oxygen radicals and by chemical reactions [21]. The challenge here is to form ordered and oriented molecular films with good electronic coupling to the solid-state surface.

### 5. Solid-state devices

Solid-state offers higher stability and wider use in electronic and optoelectronic devices. It should be noted, however, that the solid-protein interface gives rise to various interactions between the solid and the protein. The physicochemical interfacial properties, particularly hydrophobicity, and ionization, can determine the interfacial properties that can lead to surface reconstruction of the proteins [22]. Therefore, solid-state devices might not be readily available to all proteins. Among the photosynthetic proteins PSI, light harvesting protein chlorophyll complexes, and bacterial RCs were found to be sufficiently robust and most suitable for application in hybrid solid-state devices. The solid-state devices require: "better orientation of the proteins and the formation of efficient electronic junctions between the protein and the solid surface". In solid-state and electrochemical devices, the charges must be transferred through the proteins and junctions to generate efficient current. Therefore, it is important to understand the mechanisms of electron transfer through proteins and junctions.

## ELECTRON TRANSPORT THROUGH PROTEINS

Hybrid protein-solid devices were designed to explore the electrical properties of the solid-protein interactions in solid-state and in solution. It is generally accepted that proteins containing redox centers such as hemes, metal ions, quinones, and flavins acquired superior electron transport (ET) properties. The properties and the theoretical evaluation of ET across protein in solution and solid-state devices are described in several publications [4, 23-26]. Essentially, the driving force for ET is due to the difference in redox potential between the donor and acceptor. In solid-state protein-electrode junction: "the driving force for the ET process is due to the electrical potential difference between the protein and the electrodes".

**Theory of electron transport in proteins in solution**

Understanding the mechanisms of electron transport in proteins necessitated the development of a theoretical formulation encompassing the physical factors that govern these processes. Intra-protein electron transfer has been reviewed and discussed in several publications [23, 27]. Generally, electron transfer takes place by a tunneling mechanism between large atomic distances. Assuming a tunneling between two potential wells of redox centers, across a distance R in the presence of a barrier, the decay of the wave function described by Moser and Dutton is presented in Equation 1 [23]:"

$$\log_{10} k_{et} = 13 - 0.6(R - 3.6) - 3.1(\Delta G + \lambda)2/\lambda \quad \textbf{(Equation 1)}$$

where $k$ (s$^{-1}$) is the electron transfer rate, $R$ (Å) is the edge-to-edge-to-edge distance, $\Delta G$ (eV) is the driving force, and $\lambda$ (eV) is the reorganization energy. The $\Delta G$ and $\lambda$ describe the Gaussian dependence of the rate on a Marcus-like electron transfer driving force. ET of solid-state proteins placed between two electrodes differs fundamentally from ET in protein in solution.

**Theory of electron transport in proteins in solid-state**

In solid-state protein-electrode junction, the electrons are delocalized over the electrode surface. Therefore, current densities measured in solid-state experiments are orders of magnitude larger than the ET in proteins in solution [25]. Yet, as in solution, the current density is enhanced by the presence of redox cofactors in the proteins. It should also be noted that the wiring of proteins through conjugated rather than saturated molecules resulted in the formation of more efficient junctions.

Proteins in electronic devices have been modeled as one-dimensional solid-state conductors. Essentially, as described by Bostick, C *et al.* [4] the Landauer model [28] as modified in the Breit-Wigner formula can be applied to solid-state protein devices. For conductors, the Landauer model only considers elastic scattering at interfaces. The Landauer model can be extended to include molecular energy levels by applying resonant tunneling. The transmission coefficient can be modeled by an effective Lorentzian width of the molecular energy level as described by the Breit-Wigner formula.

$$(EE) = 4\Gamma_{LL}\Gamma(EE - \varepsilon\varepsilon_0)2 + (\Gamma_L + \Gamma_{RR})2 \quad \textbf{(Equation 2)}$$

In Eq. 2, $\Gamma_{LL}$ and $\Gamma_{RR}$ are the coupling energies of the molecular orbitals to the left and right electrodes, respectively, $EE$ is the energy of the tunneling electron, and $\varepsilon\varepsilon_0$ is the molecular orbital energy. This approximation is valid when $EE \approx \varepsilon\varepsilon_0$ and the separation between molecular energy levels is greater than $\Gamma_{LL} + \Gamma_{RR}$. Note that if $EE = \varepsilon\varepsilon_0$ and $\Gamma_{LL} = \Gamma_{RR}$, $(EE) = 1$, ballistic transport can be realized due to the presence of perfect resonance.

## FABRICATION AND EVALUATION OF ELECTRONIC JUNCTIONS IN SOLID-STATE DEVICES INCORPORATING PHOTOSYNTHETIC PROTEINS

The properties of the electronic junction are an important factor in the determination of the efficiency of charge transfer across the interface. Thus, covalent binding, tag-metal ion junctions, and covalent wiring of proteins to solids were found to be superior to electrostatic adsorption of proteins in the mediation of charge and current transfer. It was found that the polarity and the differences in the work function of the proteins and the solid surface are important factors in determining the direction and the energetic charge transfer. Thus, noble metals such as gold, silver, and platinum were most commonly used in electronic and optoelectronic devices. However, other metals and metal oxides such as: "aluminum, copper, zinc oxide, fluorine tin oxide, and indium tin oxide were also used in conjunction with the required work function or catalytic properties of the specific metal". In some devices, graphene, carbon nanotubes silicon, and GaAs crystals were also utilized. Affinity adsorption and covalent binding can be used for the self-assembly of protein monolayers on solid surfaces. However, oriented binding can only be achieved by binding to a solid surface through unique amino acid residues, unique tags, and specific binding sites in proteins.

The binding of proteins to solids should also facilitate the formation of efficient electrical junctions. The most common examples of junctions used in oriented binding of PSI are shown in Figure 1. Genetically engineered unique cysteine mutations in the oxidizing and reducing ends of PSI were used to form sulfide bonds with metal surface [29]. These cysteines were also indirectly bound through small molecules attached to indium tin oxide and to metal surfaces [30], and to carbon nanotubes. [31] (Figure 1a, b, 2A). Quinone derivatives such as:" 1-[15-(3-methyl-1,4-naphtoquinon-2-yl)]pentadecyl thiol bind to the vacated quinone pocket of the ubiquinone-depleted PSI and to the vacant site of Qb in PSII [32] (Figure 1c)". Another method used in direct binding of proteins involves genetically engineered histag at the N-terminal of a peptide attached to $N^{2+}$-nitrilotriacetic acid in PSI [33] and in RC [34] (Figure 1d). Poly(vinyl)imidazole Os-(bipy)$_2$Cl redox polymer binds to an electrode surface and none specifically binds to PSI at multiple sites (Figure 1e) [35] Cytochrome c attached to an electrode surface binds to the specific binding site at the oxidizing end of PSI [36, 37] and RC [38] (Figure 1F).

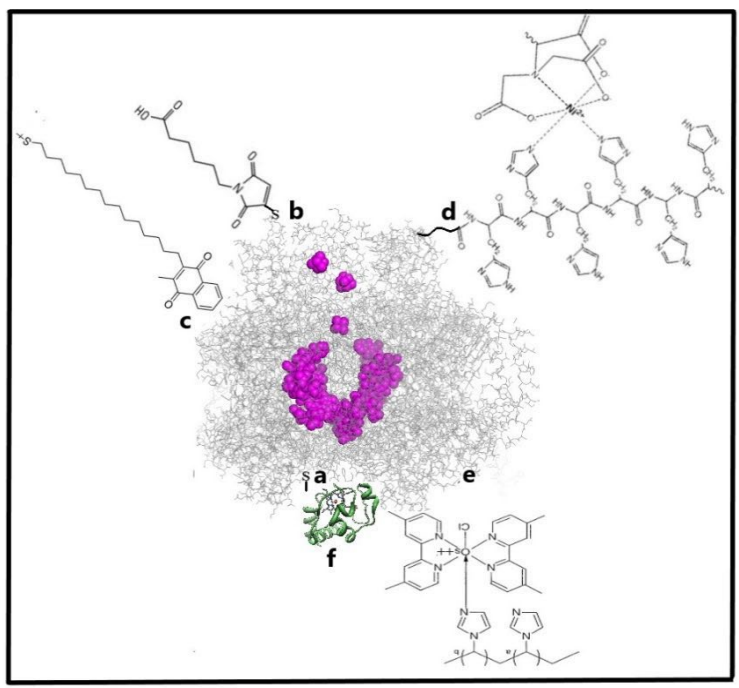

**Figure 1**. PSI-solid electronic junctions: The most commonly used electronic junction for oriented binding of PSI to solid surfaces. The junctions are also used for binding PSII and RC to solid surfaces (see text). PSI protein is represented as wire and the electron transport chain as spheres, magenta. The structure of *Synechosystis* sp. PCC 6803 PSI was taken from PDB 6hqb [39].

**Fabricating oriented PSI in solid-state devices**

Various methods were utilized for the fabrication of electronic junctions between oriented layers of PSI and electrodes in solid-state devices. Direct covalent binding of protein to solids was devised to fabricate an efficient electronic junction. The work of Frolov *et al*. [29] describes the formation of an oriented monolayer of PSI. PSI from the cyanobacterium *Synechosystis* sp. PCC 6803 was self-assembled by direct covalent binding of genetically engineered unique cysteine mutants through the formation of sulfide bonds to a gold surface. Following intensive washing and drying under nitrogen, a dense monolayer of PSI covering 60% of the total area of the gold surface was observed by atomic force microscopy (AFM) scan. All PSI molecules were functional and possessed the same polarity as determined by AFM and kelvin potential force microscopy (KPFM) scanning of the same monolayers. The photovoltage was generated by light-induced electron transport across the ~10 nm protein. PSI is a transmembrane multi-subunit protein-chlorophyll complex. The electron transfer carriers contain: "a chlorophyll *a* special pair (P700), two branches of monomeric chlorophyll *a* ($A_0$), and two phylloquinones ($A_1$) that converge on three [4Fe-4S] iron-sulfur centers (Fx, Fa, Fb) (Figure 2A)" [42-44]. Light absorbed by the antenna pigment is coherently transferred [45] to the chlorophyll *a* special pair where charge separation occurs within a picosecond. Charge separation at P700 drives electron transport across the protein generating a stable photovoltage of 1 V (Figure 3A) within ~200 ns with a quantum efficiency of ~100%.

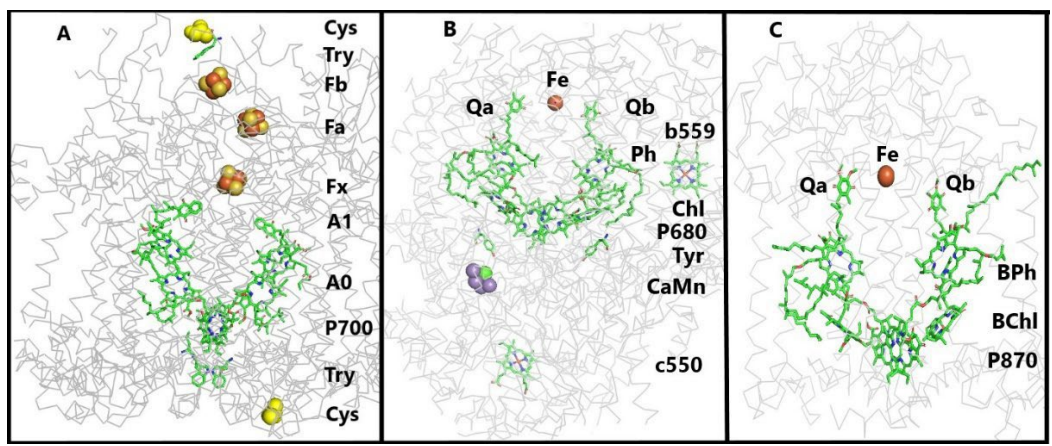

**Figure 2**: Molecular structures of reaction centers PSI, PSII, and RC. The structure presents the protein polypeptide backbone (in ribbon, gray). The molecular structure (in rods) color code (C green, O red, N blue, S yellow, Fe orang). (**A**) The electron transfer carriers of PSI contain a chlorophyll *a* special pair (P700), two branches of monomeric chlorophyll a (A0), and two phylloquinones (A1) that converge on three [4Fe-4S] iron-sulfur centers (Fx, Fa, Fb). The structure presents two cysteine mutants (yellow spheres) at the surfaces of the protein (in ribbon, gray) tryptophan molecules (Try) located between P700 and Fb at the acceptor and the donor ends of PSI, respectively. The structure of *Synechosystis* sp. PCC 6803 was taken from PDB 6hqb.(39) (B) Structure of the electron transfer carriers in PSII contains a chlorophyll a special pair (P680), two branches of monomeric chlorophyll a (Chl), two pheophytines (Ph), two plastoquinones (QA, QB), a nonheme iron (Fe), cytochrome c550 and cytochrome b559. The water oxidation site of P680 contains a Mn4Ca cluster, and two thyrosines (Tyr). The structure of *Thermosynechococcus elongatus* was taken from PDB 2axt [40]. (C) Structure of the electron transfer carriers in RC contains a bacteriochlorophyll b special pair (P870), two branches of monomeric bacteriochlorophyll b (BChl), two bachteriopheophytins (BPh) two ubiquinones (QA, QB) and a nonheme iron (Fe). The structure of *Rhodobacter sphaeroides* was taken from PDB 1aijp [41].

Various amino acids located on the exposed face of the electron acceptor side of PSI were mutated to cysteins to allow for direct thiol coupling to the Au surface. Even with the mutations being placed at an increasing distance from the P700 site, functional attachment of PSI was achieved. Although the PSI extinction coefficient is high the absorption cross-section of the ~10 nm thick single layer of PSI is low. Oriented multilayers of PSI were fabricated on Au and indium thin oxide (ITO) substrates [30]. The oriented multilayers were fabricated layer by layer by cross-linking successive layers of cysteine mutants of PSI using m-Maleimidobenzoyl-N-hydroxysulfosuccinimide ester. This was achieved by cross-linking free amine residues located on the surface of PSI monolayer to thiols located at the oxidizing side of successive PSI cysteine mutants. An increase in thickness as a function of layer number was demonstrated by ellipsometry, AFM, absorption measurements, and plasmonic effects. The multilayers demonstrated an enhanced photovoltage caused by the increase of absorption cross-section and the serial arrangement of the photosynthetic proteins. Indeed, as will be shown even extremely high photovoltage can be generated by a stack of many oriented PSI complexes. Dried micrometer-thick crystalline made of oriented plant PSI complexes generated unprecedented high photovoltages when placed on a conducting solid surface and measured by Kelvin probe force microscopy. The measured photovoltages give rise to electric fields as large as ~100 kV cm$^{-1}$, which are among the highest values ever reported in any inorganic material system [46)]. It was suggested that these high values result from the depletion of the *P700* donor centers in the PSI complex. The binding of sulfosuccinimidyl 4-(N-maleimidomethyl)cyclohexane-1-carboxylate using cabrodiinide chemistry enabled the fabrication of covalent binding of cysteine mutant of PSI to carbon nanotubes [47]. The efficient junction enabled the photoinduction of enhanced current induction in the carbon nanotubes. The induction could result from the generation of photovoltage up to 1 V produced in PSI (Figure 2A) attached to the surface of the carbon nanotubes. Formation of efficient junction was also demonstrated by covalent binding oriented chemically tagged membrane proteins [48]. Chemical tagging generally, can replace the induction of unique amino acids in organisms where genetic engineering is not available. In this process, bifunctional molecules such as 3,3´-dithiobis[sulfosuccinimidylpropionate] chemically tag the exposed surfaces of membrane proteins at selected sides of membrane vesicles. Chromatophore and thylakoid membrane vesicles containing protein chlorophyll complexes of the bacterial RC and PSI, from photosynthetic bacteria *Rhodbacter spheroide* and cyanobacteria *Thermosynechococus vulcanus*, respectively were used. After tagging the proteins were isolated from the membrane and oriented layers, which were fabricated on metal surfaces, were found functional and generated light-induced photovoltage as measured by Kalvin probe force microscopy. The polarity of the photovoltage depended on the orientation of the proteins in the layers.

**Efficiency of junctions in oriented PSI solid-state devices**

Improving solid-state devices necessitates the evaluation of the efficiency of electronic junctions formed between oriented layers of PSI and electrodes. The covalent bonding of PSI has facilitated the fabrication of efficient electronic junctions across various interfaces. This was particularly evident when a self-assembling monolayer of *N*-ε-maleimidocaproic acid was adsorbed onto n-type GaAs facilitating the covalent junction between the cysteine mutant of PSI and the semiconductor's surface [49]. Despite the distance between P700 and the solid surface being approximately 15 Å estimated from the PSI crystal structure [50], (PDB 1JB0) electrons were transmitted from the light-exited P700 to the n-type GaAs semiconductor surface at approximately 1 picoseconds. This ET rate is significantly faster than the microsecond expected for this distance, according to Equation (1). It is hypothesized that the presence of tryptophan pair W622B and W651A located between P700 and n-GaAs shortens the effective distance to 8 Å. With this adjustment and considering the energy level differences between the n-GaAs surface states and P700 ($\Delta G$ = -0.9 eV) and reorganization energy ($\lambda$ = 1 eV) an ET rate in the ps range was estimated using Equation 1. The aromatic side chains of these tryptophans, with their extensive $\pi$ electron system, serve as better electron conductors. These tryptophans that mediate electrons between cytochrome C6 and plastocyanin, electron donors to P700, support the tunneling of electrons to the solid surface. Single-molecule measurement further highlighted the effectiveness of PSI covalent binding to electrodes, demonstrating efficient electronic junction formed through the sulfide bond formation between unique cysteine mutate at PSI's reducing and oxidizing ends and gold surfaces [51] The photocurrent generated by a single PSI protein was measured using a scanning near-field optical microscope set-up, where one side of the protein was anchored to a gold surface acting as an electrode, and the other side was contacted by a gold-covered glass tip. This was made possible by using a double cysteine mutant of PSI from the cyanobacterium *Synechosystis* sp. PCC 6803 with one cysteine at the protein's surface at the oxidizing end and the second at the reducing end of the protein, enabling the formation of sulfide bonds to gold surfaces (Figure 2A). The current generated from a single PSI translated to a current density of 0.15 mA cm$^{-2}$ under 0.1 W cm$^{-2}$ illumination considering the lateral dimensions of the PSI proteins provided. The electron transfer time calculated at 16 ns suggests that electrons circumvent the terminal iron-sulfur proteins Fa/b in this configuration. Alternatively, the presence of tryptophan between the cysteine and Fb could potentially aid electron tunneling from the iron-sulfur cluster to the electrode (refer to Figure 2A). Furthermore, a tunneling mechanism may facilitate rapid electron transfer in solid-state PSI devices (see below) to a current density of 0.15 mA cm$^{-2}$ for 0.1 W cm$^{-2}$ illumination with the given lateral dimensions of the PSI proteins. The calculated electron transfer time of 16 ns suggests that the electrons bypass the terminal iron-sulfur proteins Fa/b in this setup. Alternatively, the presence of tryptophan between the cysteine and Fb might facilitate the tunneling of electrons from the iron-sulfur cluster and the electrode (Figure 2A). Yet a tunneling mechanism might facilitate rapid ET in solid-state PSI devices (see below).

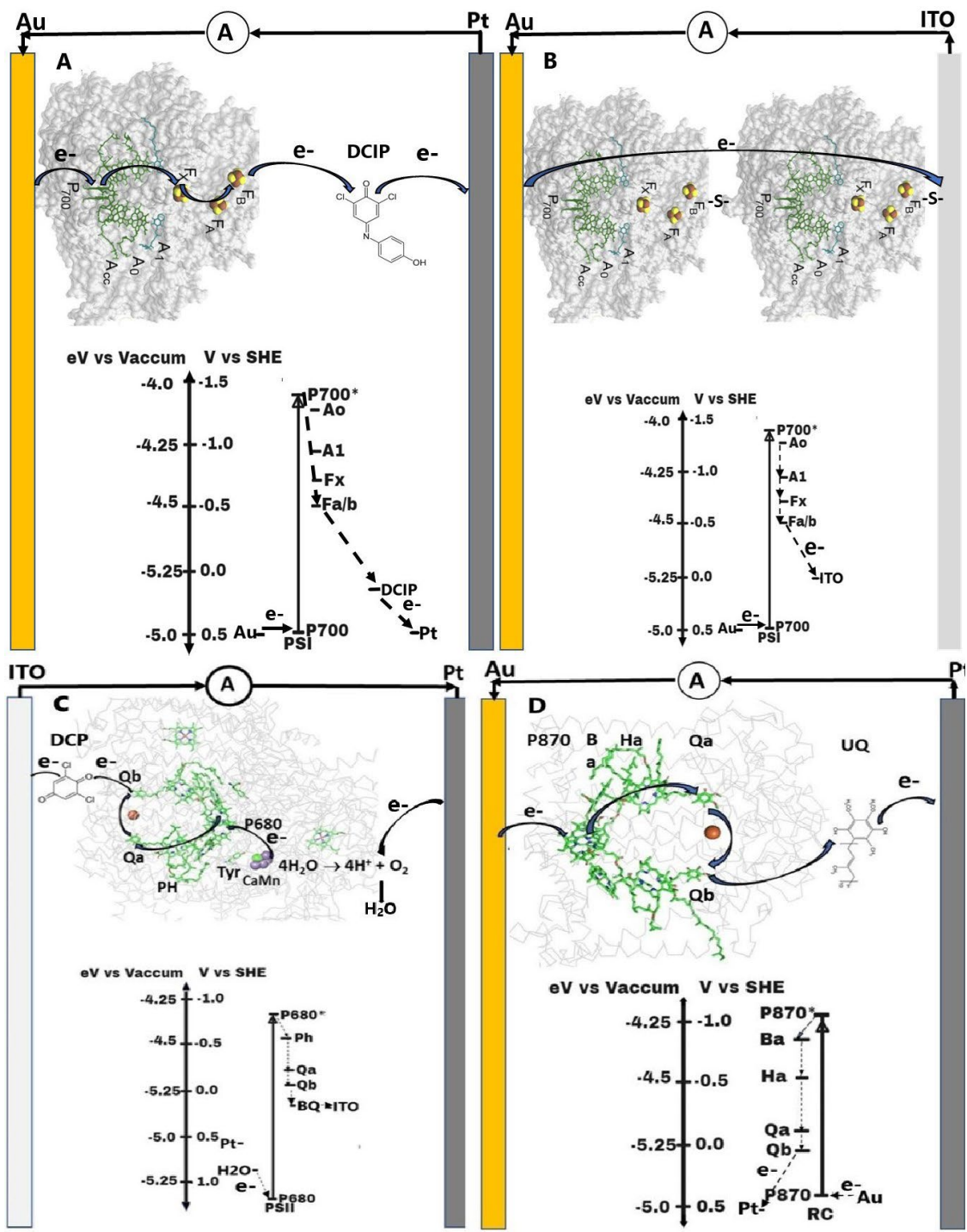

**Figure 3**: Electrochemical and solid-state devices. Examples of electrochemical devices containing (**A**) PSI, (**C**) PSII, and RC (**D**) and the energy levels of their components. An example of a solid-state device containing PSI (**B**). The schemes of the devices depict common electrode materials and mobile carriers such as ubiquinone (UQ) and dichlorophenol indophenol (DCIP). For carrier identification see the legends of Figure 2.

Electronic junctions were also engineered between metal and transparent conducting semiconductor electrodes and oriented multilayers of PSI, as depicted in Figure 3B [53]. Using photolithography, hundreds of cells in a mesh configuration, measuring 50-200 μm in width, were constructed. Due to thiol interactions, the unique cysteine mutants of PSI facilitated layer formation exclusively on the exposed surfaces of the gold and ITO electrodes. Current-voltage measurements of the solid-state cells exhibited diode-like and photodiode-like responses in the dark and under illumination, respectively. These cells demonstrated stability over several months in dry environments. An average photocurrent density of ~3.7 μA/cm² and an average open-circuit voltage of ~21 mV indicated robust electronic coupling between PSI layers and between the multilayers and the electrodes, facilitated by the formation of efficient electronic junctions. Notably, a minority of cells generated significantly higher average photocurrent densities, up to 98 µA/cm², and open-circuit voltages of 1.2 V (see Figure 4).

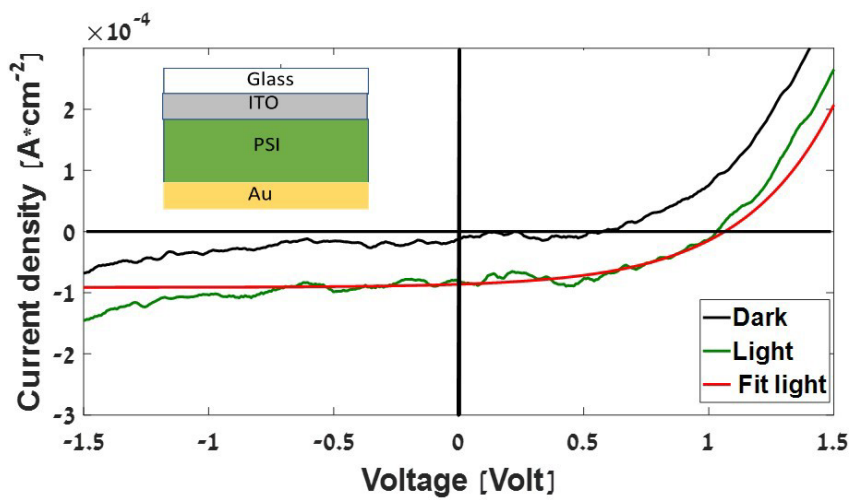

**Figure 4**: Electronic properties of the solid-state device containing PSI. Five oriented layers of cysteine mutants of PSI were covalently bound on indium tine oxide glass (top electrode) coated with N-ε-maleimidocaproic acid and topped by Au (bottom electrode). The current-voltage measurements and fit of the data to photodiode calculation were as earlier described [52]. The solid-state device layout is shown (inset). The data are taken from H. Barhom's thesis [53].

A disparity in the contact area between the electrodes and PSI multilayers could account for the significant variance in electronic properties observed between the majority and minority of the cells. This divergence may stem from variability in the contact area of one of the two electrodes deposited on the multilayers, either through spin coating of transparent conducting polymer or by evaporation of a gold layer. The current conductance likely benefitted from the presence of π-conjugated rings in Maleimidobenzoyl-N-hydroxysulfosuccinimide ester. These small molecules served to cross-link the successive oriented PSI layers in the devices. Indeed, a hopping mechanism was proposed to facilitate conductance in π-conjugated molecular wires within solid-state devices [54]. It was suggested that a tunneling mechanism aids photoconductance through PSI multilayers at the electronic junction between the multilayers and the electrodes in solid-state optoelectronic devices. A tunneling mechanism across PSI in a solid-state environment was recently proposed by Castañeda Ocampo *et al*. [55]. This suggestion was tested in a setup consisted of partially oriented self-assembled monolayers of PSI. The proteins were attached to Au substrates modified by 2- mercaptoethanol or sodium 3-mercapto-1-propanesulfonate. Current-voltage properties were measured by large-area eutectic–GaIn reaching a current density of $1.48 \times 10^{-03}$ A/cm² at an applied voltage of 1V. By varying the temperatures at which the measurements were performed, it was found that there is no measurable dependence of the current on temperature and thus, tunneling was suggested as the mechanism for transport; there are no thermally activated (e.g., *hopping*) processes. The finding of the tunneling mechanism on PSI concurred with earlier studies of ET in solid-state devices of proteins such as ferritin, azurin, heme proteins, and bacteriorhodopsin [4, 24-26)]. Such a study is represented by the current density measurement of a device fabricated by placing oriented azurin protein between a p-type silicon wafer and a gold leaf [56]. A current density of $1.5 \times 10^{-6}$ A/cm² at an applied voltage of 0.1 V across the ~3.5 nm thick azurin junction was temperature-independent. However, for both Zn-substituted and Cu-depleted azurin, thermally activated behavior was observed. These temperature-dependence differences are probably due to the function of the Cu ion. The measurements of current in solid-state proteins and junction-forming molecules are consistent with the theoretical description of tunneling as formulated in Equation 2 [25].

A distinctive characteristic of PSI electron transport is its high spin selectivity, which remains robust even under solid-state conditions [57]. This spin selectivity may play a role in enhancing conductivity efficiency. To assess spin selectivity, an oriented monolayer was constructed by forming direct sulfide bonds between unique cysteine mutants of PSI and the silver surface using dithio-bis- maleimidoethane short linkers. Spin selectivity was evaluated in a device comprising a magnet atop layers of nickel, AlOx, and silver. When nickel was magnetized such that its electron spins aligned with the polarization of PSI, electrons in the silver could tunnel between the nickel and the silver *via* the AlOx. Spin polarization was quantified as the difference in voltage measured when the Ni magnetic dipole pointed towards or away from the surface, divided by the voltage measured when no magnetic field was applied. Results revealed a spin polarization of 80%, with electrons consistently aligned parallel to their momentum. This observed spin orientation aligns with the chiral-induced spin effect observed in other chiral systems [58]. Solid-state studies have shown that coupling electron spin to linear momentum enhances electron transport by eliminating backscattering [59]. Thus, the observed spin selectivity may contribute to the efficiency of electron transfer in PSI.

**The role of photosynthetic proteins in solid-state organic-photovoltaics**
The ability of photosynthetic proteins to efficiently transfer light energy was believed to be beneficial for solid-state photovoltaic devices. Gordiichuk *et al.* [60] investigated the incorporation of PSI into organic photovoltaic devices. They deposited a partially oriented layer of PSI from the cyanobacterium *Thermosynechococcus elongatus* on the titanium oxide layer within the device. The PSI layer resulted in a substantial improvement in the electronic properties of the device reaching a short circuit current of 2.9 mA/cm$^2$, an open circuit voltage of 0.76 V, and a fill factor of 45%. However, due to the small absorption cross-section of a single partially oriented monolayer of PSI, the external quantum efficiency only reached ~0.007 % at 680 nm. In a separate experiment, Yoe *et al.* [61] incorporated a light-harvesting chlorophyll protein complex (LHCII) coupled to Ag nanoparticles into an organic photovoltaic device. It was found that incorporation of LHCII with suitable Ag nanoprisms into nano-bio hybrids device not only expands the absorption region through light trapping but also enhances the photon-absorbing activity of molecular photosensitizers through localized surface plasmon resonance from the nano-prisms. This plasmon enhancement increased the electronic output by up to 10.88%.

**Junctions in oriented PSI electrochemical devices**
Assessing the efficiency of electronic junctions between oriented PSI layers and electrodes in electrochemical devices suggests that the integration of oriented PSI layers marks a significant scientific advancement. These layers were fabricated through the formation of sulfide bonds between the thiol of unique cysteine mutant in the protein and gold electrode, as illustrated in Figure 2A. The formation of these oriented multilayers was facilitated by a small platinized metal junction deposited at the reducing end of PSI by photoreduction of platinum ions. The thiol moiety of the of the cysteine mutants of PSI located at oxidizing end was sequentially attached to the platinized reducing end yielding an electronically coupled oriented multilayer [62]. Notably, the photovoltage exhibited an increase of up to 2.4 folds from single to three layers, accompanied by a rise in the photocurrent up to 120 mA in an electrochemical cell containing 2,6-Dichloroindophenol as an electron carrier. The A

The assembly of an oriented PSI monolayer on a gold surface was achieved through the utilization of a recombinant poly-histidine tag. Specifically, recombinant PsaD-His6 was immobilized on Ni$^{2+}$-nitrilotriacetic acid chelation moiety functionalized on thin Au layered on glass-coated ITO surface [63]. Subsequently, this assembly was exposed to native PSI complexes, whereby the intrinsic PsaD subunit naturally exchanged at 65% efficiency with an excess of the recombinant PsaD-His6. This exchange resulted in the immobilization of PSI with P700 facing away from the ITO/Au transparent electrode. The electron transport from PSI to the Ag electrode was mediated by tris(8-hydroxyquinoline) aluminum. Similarly, a solid-state device was fabricated by binding a photosynthetic bacterial reaction center that was modified by the formation of histag at the M subunit of the protein. The calculated internal energy conversion efficiency of 12% attests to the viability and efficiency of the hostage electronic junctions in this device. However, it is noteworthy that the external efficiency of a device containing a monolayer of proteins is constrained by the low light absorption cross-section. Nevertheless, the surface area of the monolayers was dramatically increased by self-absorption of tag PSI on electrodes constructed from tall ZnO nanowires grown on ITO-glass and 60 nm-pore TiO2 nanocrystalline photoanodes [33]. PSI was assembled by exchange of the native subunit with immobilized E-subunit bound to TiO$_2$ by histag and to the ZnO nanowires through the ZnO tag. The uneven surface necessitated the mediation of mobile electron carrier Co(II)/Co(III) in the ion-containing electrolyte to transfer current from the cathode to the proteins in the electrochemical devices. Remarkably, under standard sunlight, the devices yielded record values of photocurrent density of 362 mA/cm$^2$, open circuit photovoltage of 0.5 V, fill factor of 71%, and electrical power density of 81 mW/cm$^2$. Nevertheless, the total incident-light to electrical external power conversion efficiency was only ~0.07%.

Another approach for the formation of an oriented layer consisted of the formation of a covalent bond between platinized PSI and modified electrode. PSI with a redox potential of –0.53 V enabled the photoreduction of Pt$^{4+}$ and the deposition of metallic platinum at the reducing end of PSI [64]. PSI and glucose oxidase were assembled to fabricate photo-bioelectrochemical electrodes. The platinized PSI complexes from cyanobacterium *Mastigocladus laminosus* were attached to modified indium tin oxide. PSI surface amines were then connected to the pyrroloquinoline by a similar reaction yielding partially oriented monolayers. The cathode consisted of a redox polymer attached to a glucose oxidase assembly. A photocurrent of up to 0.4 μA (bias voltage -0.05 V) was generated by the photochemical cell [65]. Another unique method for oriented binding and formation of the electronic junction consists of wiring quinone-depleted PSI through the naphthoquinone derivative attached to an Au electrode [66]. PSI from the thermophilic cyanobacterium, *Thermosynechococcus elongatus*, was connected to an Au microelectrode, by a self-assembled PSI attached to Au-nanoparticles. The Au-nanoparticles were functionalized with a thiolated naphthoquinone derivative and attached to the Au electrode through dithiol small molecules. The naphthoquinone derivatives were bound to the vacated quinone pocket of the ubiquinone-depleted PSI. The efficiency of the electronic junction was demonstrated in electrochemical measurements.

Enhancing the binding orientation of PSI by the formation of efficient wiring to the electrodes mitigated the interference from unoriented complexes and amplified the efficiency of the generated photocurrent. This orientation was obtained through covalent bonding between PSI and the electrode. The binding was fabricated either directly or through a mediated tedder. The assembly of oriented multilayers has proven to augment the output per unit area. However, the optimal outcome was attained by expanding electrode surface area, accomplished through the construction of elongated nanowires employing porous ITO substrate.

**Efficiency of junctions in partially oriented PSI electrochemical devices**
Initially, devices containing partially oriented PSI proteins were designed for the research and the development of electrochemical experiments. However, the ready availability and easier procedure of fabrication yielded a wealth of important scientific information. Thus, only partially oriented layers were obtained by affinity binding of PSI on gold electrode coated with self-assembled monolayer containing head groups of \OH, \COOH. It was found that for the self-assembled monolayer of mercaptoacetic acid monolayer, 83% of the electron transport vectors were parallel to the surface, whereas in the 2-mercaptoethanol monolayer, 70% were oriented perpendicular to the surface, as determined by scanning tunneling microscopy. The self-assembled layer for PSI adsorption at the P700 end allows for the electron transfer from the electrode to the electron acceptor methyl viologen. The presence of PSI in photoelectrochemical experiments caused an enhancement of current from the Au-coated electrode, indicating that at least some of the proteins were attached to the electrodes at the P700 end of PSI [67]. The problem of lack of fabrication of a fully oriented layer by adsorption of PSI to a modified electrode in an electrochemical experiment was solved by the use of a mobile electron mediator that could be oxidized and reduced by PSI and by the cathode and anode of the device. High values of photocurrent were generated by the use of ferricyanide as a mobile electron mediator from a partially oriented mass of PSI deposited on an electrode [68].

A more specific junction was achieved by utilizing cytochrome c as a mediator of electrons between PSI and an electrode. Cytochrome *c* is the natural electron donor that binds to a specific site in PSI and therefore can form oriented binding of the reaction center to an electrode. A light-activated electron-transfer chain was assembled using PSI from *Thermosynechococcus elongatus* as the photoactive enzyme, cytochrome c6 as the electron donor, and methyl viologen as the electron acceptor [36]. Under steady-state conditions second-order rate constants of $5.9 \pm 0.1 \times 10^6$ M-1 s-1 were determined for electron transfer from reduced cyt c6 to photooxidized PSI and $9 \pm 0.1 \times 10^6$ M-1 s-1 from PSI to methyl viologen.

An interesting approach for the formation of an electronic junction involves utilizing a reducing polymer to bind PSI and facilitate current mediation between the reaction center and the electrode. Multilayers of partially oriented plant PSI were assembled on gold electrodes coated with the transparent conducting polymer poly(3,4-ethylenedioxythiophene)-polystyrenesulfonate [69]. The deposition process consisted of alternating between layers of polymer and PSI, and up to 9-layer pairs of PSI and polymer were obtained. When tested in an electrochemical cell employing ubiquinone-0 as a mediator, the photocurrent performance of 6-layer pair samples yielded a photocurrent of $414 \pm 13$ nA/cm$^2$. However, the efficiency of the electronic junction remained a challenge. Numerous efforts have been made to identify the most effective material for creating efficient junctions with PSI. Graphene has gained interest in material chemistry due to its high conductivity, transparency, mechanical strength, and lightweight. These properties render it a superior material for electrode fabrication in opto-electrochemical devices. In a novel approach, a hybrid graphene-PSI electrode was fabricated by absorbing PSI proteins extracted from spinach as a monolayer onto graphene *via* vacuum-assisted casting. In the absence of a covalent junction, a mobile carrier methylene blue was used to shuttle electrons between PSI and the electrodes within the electrochemical cell, generating a photocurrent of 0.55 μA/cm$^2$ from a monolayer of PSI on graphene [70]. In another attempt, the π-π interactions between naphthalene derivatives and graphene electrodes facilitated the fabrication of an electronic junction with the adsorbed PSI [71]. In a partially oriented monolayer of PSI adsorbed on the naphthalene derivative-modified graphene electrode only the proteins oriented with the acceptor end attached to the electrode surface could generate photocurrent *via* soluble electron acceptor, methyl viologen. By applying an overpotential and addition of a mobile carrier the photocurrent density can be further magnified to $20 \pm 0.5$ μA/cm$^2$. Another approach to forming a junction between graphene and PSI involves spin-coating the graphene electrode with bipyridine-modified graphene oxide particles. The particles served as a matrix for the adsorption of partially oriented PSI [72]. It was found that the addition of cytochrome c facilitated additional electronic junction between the oxidized graphene and PSI, resulting in photocurrents up to ~14 µA/cm$^2$, corresponding to a turnover frequency of 30 e$^-$ PSI$^{-1}$ s$^{-1}$ and external quantum efficiency of 0.07% at a thickness of about 15 µm.

An electronic junction of partially oriented PSI was also fabricated by coating a gold electrode with a redox polymer [8]. Isolated PSI from the cyanobacterium Thermosynechococcus elongatus was immobilized on a gold electrode surface via an Os complex-containing redox polymer, poly(ethylene glycol)diglycidyl ether, which served both as an immobilization matrix and as the electron donor for PSI. Subsequently upon addition of methyl viologen as a sacrificial electron acceptor, a catalytic photocurrent with densities of up to 29 µA cm$^{-2}$ at a light intensity of 1.8 mW cm$^{-2}$ was observed upon illumination. An electrochemical device was also constructed, consisting of redox hydrogel film connected between PSI and an electrode for photocurrent generation [35]. A poly(vinyl)imidazole Os(bispyridine)2Cl polymer redox hydrogel film was used for achieving maximum electron-transfer rates from a glass carbon electrode to attached PSI complexes from *Thermosynechococcus elongatus*. The PSI-containing redox hydrogel film displayed electron-transfer rates of up to 335 e$^-$ s$^{-1}$ PSI$^{-1}$, significantly surpassing the rates observed in natural photosynthesis. Under O$_2$ supersaturation, in the presence of methyl viologen as a mobile electron mediator, a photocurrent of 322 µA cm$^{-2}$ was achieved. A drawback of the electrochemical device is its relative instability under continuous illumination in the presence of oxygen. However, the use of 2,3- dimethoxy-5-methyl-1,4-benzoquinone, as an electron acceptor for the PSI-based photocathode, extended the lifetime of the device [73]. Moreover, under the exclusion of O$_2$ even after 15 h of continuous operation, 26% of the initial photocurrent was preserved.

The biophotovoltaic cell was further enhanced with an assembly comprising a PSI-based biophotoanode and an oxygen-reducing glucose oxidase/peroxidase cathode. [74] The Au anode was modified with the redox polymer poly(3-isopropyl methacrylate-co-butyl acrylate-co-glycidyl methacrylate)-viologen to which Longmuir monolayer of PSI from Thermosynechococcus elongatus BP-1 was attached. The PSI layer was then coated by poly(1-vinylimidazole-co-allylamine)-[Os(2,2'- bipyridine)2Cl]Cl, embedding glucose oxidase. The cathode consisted of a microporous surface of the carbon cloth material. Peroxidase and glucose oxidase enzymes were modified with 1-Pyrenebutyric acid N-hydroxysuccinimide ester and attached to the electrode through a π-π between the pyrene moiety and the carbon electrode. Glucose oxidase at the anode provided electrons to the photo-oxidized PSI by oxidizing glucose to gluconolactone. Glucose oxidase at the cathode oxidized glucose, producing oxygen peroxide that was reduced to H$_2$O by peroxidase-accepting electrons from the cathode. The cell achieved "a maximum current density of ~7.2 µA cm$^{-2}$ and exhibited a maximum power output of ~3.7 µW cm$^{-2}$ at a voltage difference of 600 mV. The biophotoelectrode exhibited reasonable stability, retaining 65% of the initial photocurrent response after 90 min of the experiment albeit requiring the consumption of glucose".

**Efficiency of junctions in partially oriented PSII electrochemical devices**
Analyzing electronic junctions between partially orientated PSII complexes and electrodes in electrochemical devices reveals that adsorption of PSII on the electrode is required for the transfer of the photoelectrons from the final electron acceptors QA and QB to the electrode (Figure 2C) [4, 43]. In one of the early attempts partially oriented cyanobacterial PSII was self-assembled and was first adsorbed as monolayer phosphonic acid on mesoITO [75]. The negatively charged carboxylate attracts the positive dipole on the electron outlet side of PSII *via* Coulomb interactions allowing direct photoelectron transfer to the electrode in an electrochemical device. Covalent binding of the carboxyls to the PSII amines enhances the anodic photocurrent. However, the electronic junction between the electrode and the partially oriented proteins yielded low photocurrent. The photocurrent increased tenfold to 4.5 µA cm-2 in addition to the mobile electron mediator. Yet, the system was unstable because of photodegradation caused by photoexcited chlorophyll *a* triplet states undergoing energy exchange with 3O$_2$ to give deleterious 1O$_2$ and other reactive oxygen radicals in the electrolyte solution. PSII was also adsorbed on an increased surface area of an electrode by the fabrication of "three-dimensional opal–indium tin oxide and graphene electrodes [76]. In the presence of electron mobile carrier 2,5-dichloro-1,4-benzoquinone a photocurrent of ~200 µA/cm-2 was obtained".

The addition of hydrogenase to the cathode of an electrochemical device improved the photocurrent generated by PSII attached to the anode. PSII dimers isolated from *Thermosynechococcus elongatus* were adsorbed on an ITO anode with a macroporous inverse opal architecture and a mesoporous skeleton [77]. The mesoporous ITO skeleton increased the effective surface area for PSII interactions and facilitated electronic coupling. The electrochemical device contained an inverse opal- mesoITO H$_2$-ase cathode. Photocurrent density of up to 450 ± 10 µA cm$^{-2}$ was observed at 0.9 V (illumination at 660 nm10 mW cm$^{-2}$). These setups enabled to sustain the activity of PSII for almost an hour.

Another strategy to enhance the loading of PSII onto electrodes is to entrap PSII in a redox-active polymer matrix on an electrode surface. In this approach, PSII of any orientation can in principle be efficiently connected to the electrode by the redox-active moieties that are homogeneously distributed in the matrix, which can mediate charge transfer via an electron hopping mechanism [78]. An effective enzyme immobilization strategy involves the use of a structured electrode surface to increase the available surface area. PSII was adsorbed on a structured indium tin oxide (ITO) electrode that incorporated macroporosity and mesoporosity with high thickness. As a result, a 1600-fold increase in PSII loading was observed compared to conventional flat electrodes. The use of a phenothiazine-modified redox polymer matrix to enable efficient mediation of PSII photoelectrons to the electrodes yielded maximum photocurrent values of 577 µA cm$^{-2}$ and external quantum efficiency of 9.3% under illumination with monochromatic light at 685 nm at the absorption maximum of PSII. Although relatively stable, a loss of 83% of the initial activity was recorded after 60 min of illumination.

**Efficiency of junctions in oriented PSII electrochemical devices**
Exploring electronic junctions between oriented PSII layers and electrodes in electrochemical devices reveals an enhanced efficiency in the electronic junctions between PSII and the electrodes through the orientation of PSII binding to the electrodes. One approach involved binding through histag, while the other utilized wiring of PSII through quinone. In the first method PSII was attached to the electrode by genetically introducing histag at the CP47 subunit from the cyanobacterium *Thermosynechococcus elongatus*. PSII was attached to gold electrodes modified with thiolates bearing terminal N$^{2+}$-nitrilotriacetic acid groups [79]. Since PSII was attached to the electrode at the water oxidation side, a mobile electron carrier, 2,6-dichloro-1,4-benzoquino, was employed to shuttle electrons from QA to a platinum electrode. Upon illumination, a photocurrent density of approximately 14 A/ cm2 (bias +0.3 V) was generated for oxidation of the mobile carrier. The second approach involved wiring of PSII through quinones. PSII/benzoquinone polymer-based photo-bioelectrochemical cells were fabricated using PSII isolated from the thermophilic cyanobacterium *Mastigocladus laminosus* [32]. Initially, the photoanode was prepared by the electropolymerization of mercapto benzoquinone on an Au surface. PSII was then deposited onto the quinone layer with the intention that the quinone units would bind to the QB site of PSII (Figure 2C). An electrically wired bilirubin oxidase/carbon nanotube-modified cathode was employed in this setup, resulting in the high turnover rate of 518 e-s-$^1$ between PSII and the electrode. Voloshin R. A. *et al*. [80] have presented a description of the multiple efforts to construct PSII-based electrochemical devices. The review also describes semiartificial PSII-based photochemical devices which were not addressed in this review.

**Development of electrochemical devices utilizing PSI and PSII composite**
Electrochemical devices containing PSI and PSII were constructed to mimic the natural z system in thylakoids. Utilization of both photosystems energized by light was expected to be advantageous. Layered assemblies of "PSI and PSII from *Thermosynechococcus elongatus* on ITO electrodes were constructed via a layer-by-layer deposition process [81]. Poly N, N ′-dibenzyl-4,4′-bipyridinium (poly-benzyl viologen) was employed to mediate electrons from PSI to the ITO electrode, while the redox polymer polylysine benzoquinone mediated electron transfer from PSII to PSI. Illumination generated a photocurrent of ~1.2 µA/ cm$^2$. The study indicated that the transfer of electrons from PSII to PSI through polylysine benzoquinone was the rate-limiting step in the process".

An electrochemical system that contained a PSI-based photocathode and a PSII-based photoanode resulted in the fabrication of a photovoltaic cell functioning as a closed system [35]. This setup comprised of two compartments connected by a salt bridge. In one cell, gold disk electrodes coated with PSI from *Thermosynechococcus elongatus* were integrated into the redox hydrogel Os2 [35]. In the second cell, PSII was integrated into a gold electrode modified by the redox hydrogel Os1 [8]. Under illumination: "Water was oxidized to oxygen in the anodic compartment by PSII, while in the cathodic compartment, oxygen was reduced by PSI, *via* methyl viologen. The cells generated a short-circuit current of ~2 µA/cm$^2$, and an open-circuit voltage of ~90 mV. The application of protective polymer membrane-alike layers seems to decelerate the gradual degradation in the activity of the photosynthetic active units by serving as mechanical shields".

**Efficiency of junctions in oriented RC devices**
Assessment of the efficiency of electronic junctions between oriented reaction centers (RC) from photosynthetic bacteria and electrodes could advance the research in this scientific field. The bacterial reaction center consists of only three protein subunits, four bacteriochlorophylls, two pheophytins, and one carotenoid molecule. The RC was expected to be a good candidate for an active component in photochemical cells because of its relatively simple structure and its absorption that extends to the ~800 nm region. Electrochemical devices containing RC require attachment of the protein to a working electrode and transfer of the photo-generated electrons from Qa or Qb [43] to the cathode (Figure 3D). Genetically engineered histag attached to the terminal amino acid of the H subunit of the intact RCs from *Rhodobacter sphaeroides* enabled the fabrication of an oriented monolayer of the protein on a solid surface [82]. The histag-RCs were immobilized on an indium tin oxide electrode overlaid with nickel-nitrilotriacetic acid on a gold substrate. An opposite polarity of an oriented monolayer was obtained by the use of genetically engineered histag at the M-subunit of RCs from *Rhodobacter sphaeroides* [83]. The histag-RCS were immobilized on nickel- nitrilotriacetic acid-modified gold electrode. In the presence of an electron acceptor, ubiquinone-10, illumination of this RC electrode generates a cathodic photocurrent in an electrochemical device (Figure 3D). Photocurrent density was further improved by the addition of cytochrome c to the oriented monolayer of histag-RCS [84]. Cytochrome c acts as a diffusible ET mediator to the RC primary donor. Precisely adapted by evolution for the interaction with the RC, cytochrome c penetrates inside the RC protein at the side of the primary donor (special pair, P) leading to a high efficiency of ET between these two proteins [85]. Substantial improvement in the efficiency of an electrochemical device was achieved by oriented binding of histag-RCs to carbon nanotubes [34]. For protein immobilization on carbon substrates, a bifunctional linker having a pyrene group at one end for the attachment to the electrode and a nickel-nitrilotriacetic acid group at the other end for the attachment to the histag-RCs was used. The $Al_2O_3$ spacer located between the carbon nanotubes prevented the binding of RC on the outside and allowed the self-assembly of RCs inside the nanotubes. An efficient electronic junction was formed by the addition of cytochrome c which binds to the protein and mediates electrons between the electrode and the primary donor P870 at the acceptor end of the RC. Ubiquinone-2 served to shuttle electrons from RCs to a platinum anode. The photocurrent density increased from 0.314 $\mu A/cm^2$ to 1.414 $\mu A/cm^2$ (electrodes potential 50 mV, light intensity 5 mW·$cm^{-2}$) when RCs were bound to the surface of the graphite electrode and the carbon nanotubes layered on graphite electrode, respectively. The results obtained in the present work demonstrate that the attachment of RC to the inner carbon nanotube walls increased the electron transfer rate between RC and electrode and improved protein packing to higher functional density and better light harvesting efficiency.

The RC is embedded in the membranes of the bacteria. It was therefore important to verify whether adding a lipid bilayer might improve the electrochemical properties of oriented RCs in these devices [86]. In these experiments: "RCs from *Rb. Sphaeroides* with a genetically engineered 7-his-tag were bound to a Ni-NTA-modified gold surface. Subsequently, the bound RCs were subjected to lipid micelles to form lipid-tattered proteins. Photocurrents were generated in the range of 10 $\mu A\ cm^{-2}$, however, different from previous studies at potentials of −200 and −300 mV, without cytochrome c as a mediator. The unexpected behavior is explained by an inter-protein reaction between RC molecules promoted by the lipid bilayer."

Cytochrome c was also utilized as a junction to a modified gold electrode and for partial orientation of the RC monolayer [38]. A single cytochrome c-RC junction might be a rate-limiting step in the photocurrent density. Indeed, improved photocurrent was obtained in a novel setup consisting of RCs attached directly to a modified gold electrode *via* hydrophobic interactions and a film of six cytochrome c per RC electrostatically bound to the electrode [87]. This resulted in larger photocurrent onset potentials, positively shifted half-wave reduction potentials, and higher photocurrent densities reaching 100 $\mu A\ cm^{-2}$.

It was found that "an improved device of biophotoelectrodes comprising an IO-mITO matrix suffused with purple bacterial RC-LH1 proteins and cytochrome c reached steady-state photocurrents of around 0.25-0.3 mA $cm^{-2}$ [88]. The steady-state photocurrent was sustained for many hours by supplementing cytochrome c which stabilized the electrical connection between the photoactive RC- LH1 complexes and the IO-mITO electrode." These values are still lower than the photocurrent generated by PSI and PSII electrochemical devices. A simple device contained chromatophores from the photosynthetic bacterium *Rhodobacter sphaeroides* that were attached to a Millipore membrane filter. In these experiments the membrane was sandwiched between two semiconductor indium tin oxide [89]. At ~26 mV the photocurrent reached a value of ~200 $nA/cm^2$ in the presence of mobile electron transfer mediators and the inhibitor antimycin A, an inhibitor of cytochrome *bc*1 complex.

**PHOTOSENSORS EMPLOYING PSI AND RC IN THE GATES OF FIELD EFFECT TRANSISTORS**
PSI demonstrates significant promise for application in semiconductor technology owing to its exceptional electro-optical properties, cost-effectiveness, and low thermal energy. The incorporation of PSI in an integrated circuit is anticipated to yield a high-density and sensitive photosensor, capable of detecting a single photon at room temperature, particularly through its incorporation in a field effect transistor (FET). Notably, a sensing device utilizing PSI from *Thermosynechococcus elongatus* was successfully fabricated with electric charges converted to electric signals by an ion-sensitive field-effect transistor [90]. Essentially, a high-density, low-power source-drain follower was engineered for a sensor array, achieved by depositing Au/Ti extended-gate electrodes on standard metal-oxide semiconductor field effect transistor, followed by deposition of thin silicon nitride layers *via* catalytic chemical vapor deposition. Partially oriented PSI was electrostatically attached to an Au electrode coated by 3-mercapto-1-propanesulfonic acid within electrochemical cells, yielding a cell output voltage of 120 $\mu V/\mu Wcm^{-2}$, and resulting in the production of images of reasonable quality. In another study, an enhanced photosensor employing PSI within an electrochemical gate of a field-effect transistor was constructed [91]. PSI tethered to a gold nanoparticle was attached to the gate surface within a small electrochemical cell.

A molecular wire was constructed by reconstitution of PSI from the cyanobacteria *Thermosynechococcus elongatuswhich,* which had been depleted from the quinones in the electron transfer chain. This quinone-free PSI was reconstituted with 1 naphthoquinone derivative tethered to gold nanoparticles, forming a molecular wire. Subsequently: "the reconstituted PSI was immobilized on the gate of a field effect transistor *via* silane coupling reaction and Au-S bond formation. Notably, light irradiation induced a significant change in voltage between the gate and the source, from -3.3 V to -5.4 V, when the current between the drain and the source was retained below 1 μA". The output of an image projected onto the gate was processed by a computer at a resolution of 60×80 pixels. This work presents a practical approach for the establishment of an efficient junction between PSI and a transistor utilizing a molecular wire, thus establishing the fabrication of a sensitive bio-photosensor.

A photosensor utilizing RC in a gate within an electrochemical field effect transistor was constructed [92]. The RC derived from *Rhodobacter sphaeroides* was affixed to the gate's surface *via* a monolayer of cytochrome c (Figure 1f). Initially, a self-assembled monolayer of cytochrome c was attached through a linker to the transistor. Subsequently, a monolayer and RC were attached *via* the cytochrome binding site of the protein. Approximately, 33% of the cytochrome c molecules on the transistor gate effectively couple with RCs. Upon exposure to light irradiation, a notable change in the voltage between the gate and the source was observed. Under a gate-source voltage of 3 V, the saturation current at a drain-source voltage of 1.2 V increased from 34 mA for the transistor lacking cytochrome c to 71 μA. Due to the heightened sensitivity of the transistor, this method can be applied to fabricate photosensors capable of detecting low light intensities.

**LIGHT-MATTER INTERACTIONS**

Hybrid light-matter states are generated through the interaction between molecular excitons and plasmons. The light-matter states will occur when the resonance of the plasmons is tuned with the energy of molecular transitions. Strong coupling represents an intensified form of light-matter interaction, leading to the creation of new polaritons. Hybrid light-matter states form when the exchange of energy between the resonant components surpasses any decay process. These two new hybrid states possess energy levels distinct from plasmons and polaritons. In strong coupling, the resulting energy levels are referred to as vacuum Rabi splitting [1, 2]. The formulation of Rabi splitting is presented in Equation 5:

$$\hbar\Omega_R = 2V_n = 2\mathbf{d}\cdot\mathbf{E}_0 = 2d\sqrt{\frac{\hbar\omega}{2\varepsilon_0 v}} \times \sqrt{n_{ph}+1} \qquad \textbf{(Equation 5)}$$

where $\hbar\Omega_R$ is the Rabai slitting, $V$ is the interaction energy between the electric component of the electromagnetic field in the cavity, $E_0$ is the transition dipole moment of the material, $d$, $\hbar\Omega$ is the resonance energy, $\epsilon_0$ the vacuum permittivity, $v$ the mode volume and $n_{ph}$ the number of photons present in the system. Even when $n_{ph}$ goes to zero, $\hbar\Omega_R$ has a finite value. Under strong coupling, the Rabi splitting can be as large as ~1 eV, corresponding to a significant fraction of the original transition energy.

Strong coupling causes modifications of the ground-state properties. Indeed, it has been found that strong coupling induces changes in the work function of an organic material" [3].

Surface plasmons have been found to act as resonators, generating strong coupling. Consequently, strong coupling can not only modify the spectral response of absorbers but also effectively direct energy to specific molecules or detectors. This phenomenon was exemplified by the demonstration of strong coupling in single molecules of methylene blue placed within plasmonic nanocavities [93]. Furthermore, strong coupling has been shown to enhance metal-to-semiconductor electron transfer [94]. In cases where the interaction between plasmons and the molecules is weak, surface plasmon influences physical phenomena without perturbation of the wave functions. In weak coupling: "surface plasmon resonance arises from the collective oscillation induced by the electromagnetic field of light. This can be utilized to investigate the interaction between light and matter beyond the diffraction limit [2]. The weak coupling has been observed to induce various light-matter interactions, including surface-enhanced Raman scattering [95] surface plasmon-enhanced absorption [96], enhanced fluorescence [97], and fluorescence quenching. The study of weak coupling has contributed to various applications such as light-emitting diodes, solar cells, low-threshold lasers, biomedical detection, and quantum information processing. Several photophysical phenomena are known to be enhanced by surface plasmons, including surface-enhanced Raman scattering, [98, 99] surface-enhanced infrared absorption, [100] and surface-enhanced fluorescence [101]. Additionally, investigations have explored the enhancement of absorption in photovoltaic systems by plasmonic metal nanostructures [99, 102, 103]. It has also been shown that: "the Förster-type nonradiative transfer rate can be increased by a factor of 7 with an efficiency approaching unity" [104].

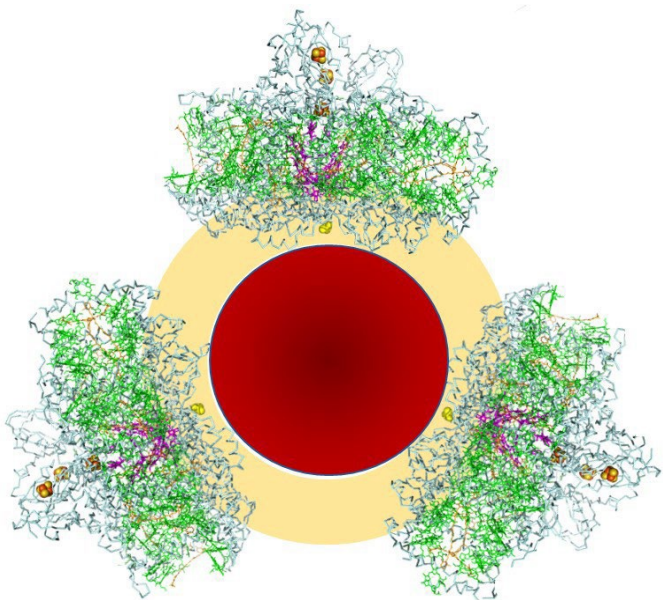

**Figure 5**. PSI attached to gold nano-particle. PSI was covalently attached to the gold nanoparticles (~10 nm) through the formation of a sulfide bond between the cysteine thiol and the metal. For details see Carmeli, *et al*. [105].

**Plasmon-enhanced absorption and photocurrent in PSI devices**
Theoretical prediction suggested an enhancement of the absorption of the PSI facilitated by metal nanoparticles [106]. It was assumed that in metal nanocrystal-PSI hybrids, both the optical absorption of the antenna chlorophylls and the rate of photogenerated electron transfer could be enhanced. These chlorophylls absorb photons and transfer optically created excitons toward the reaction center. The efficient light harvesting stems from the fact that the Förster transfer time is shorter than the transfer time to the metal. The photogenerated electron and hole rapidly separated spatially, occurring within about 1 ps. This time is shorter than that required to transfer exciton energy to a metal nanocrystal. The enhancement factors of photogenerated electron transport for the gold and silver nanocrystals (radios 10nm) range from 5 to 15. Metal nanoparticles can function as artificial antennas augmenting the light absorption of the protein. Experimental validation of this concept involved a nano-particle-PSI hybrid system (Figure 5) which exhibited an increase in light absorption and circular dichroism across the protein's entire absorption band, rather than at the specific plasmon resonance wavelength of spherical metal nanoparticles [105, 106]. For the Ag-PSI system, an approximately 2 to 4 enhancement factors were measured for the 680 nm peak. In the case of the Au-PSI system, enhancement of approximately 1.9-2.5-fold was observed in light absorption for the 440 and 680 nm bands.

The impact of plasmon resonance on photocurrent was investigated in a separate experiment. Oriented assembly of red alga PSI complexes from *Cyanidioschyzon merolae* on a silver island film induced strong enhancement of both the fluorescence intensity and photocurrent [37]. The electrochemical device comprised of a glass electrode coated with flour tine oxide onto which silver islands were deposited. The PSI complexes were oriented on graphene through Co-NTA-pyrene and His6-tagged cytochrome $c_{553}$, positioned on top of the silver islands. In the absence of an external bias, the photocurrent output showed a 22.6-fold improvement compared to PSI without silver islands. At an applied voltage of -300 mV, the current density output was 224.4 nA cm$^{-2}$. The relatively low output is probably due to the rather long (70 Å) junction and low density of PSI coverage of the electrodes.

Plasmon-induced photocurrent enhancements were also recorded for PSI immobilized on Fischer patterns of silver nanopyramids [107]. The plasmon-enhanced photocurrents indicate enhancement factors of ~6.5 and ~5.8 as compared to PSI assembly on planar Ag substrates for nominal excitation wavelengths of 660 and 470 nm, respectively. As was earlier theoretically calculated [106] plasmon-induced photocurrent enhancements of ~6.8 and 17.5 were generated in highly ordered Au and Ag nanodisks [108]. The enhancement was induced by dipolar plasmon resonances tuned to ~680 and 560 nm, respectively, compared to PSI assembled on planar ITO substrates. The Au and Ag nano-disc arrays with a 200 nm pitch were patterned *via* electron beam lithography on ITO glass coated with a self-assembled monolayer of OH-terminated thiol. A partially oriented monolayer of PSI was adsorbed between the Au and Ag nano-disc arrays. Although the photocurrent density of the PSI monolayer was only ~30 nA/cm$^2$ the plasmon enhancement was as high as expected from the theoretical calculations. Enhancement of photocurrent by PSI immobilized on a gold nano-slit electrode has also been demonstrated. [109]. The electrochemical device consisted of a gold electrode on a glass substrate in slits ranging from 50 to 400 nm, fabricated by photolithography and ion etching. The gold slits were chemisorbed by a monolayer of mercaptoundecanoic that covalently binds to free ammine residues of the proteins, forming a partially oriented self-assembled monolayer of PSI. Methylene blue in electrolyte solution was used to shuttle electrons between PSI-coated slits and a second electrode. The photocurrent varied according to the slit size, reaching a maximum of 9.95 nA/μm$^2$ for a width of 100 nm. There was a good correlation between the surface plasmon generation efficiency of nano-slits and the photocurrent obtained experimentally by immobilizing PSI on the gold nano-slit electrode surfaces with different nano-slit widths. These results could be attributed to the phenomenon of plasmon-exciton coupling effect on the PSI in the nano-slits. The energy conversion to photocurrent is greatly enhanced, in correlation to the surface plasmon generation efficiency and electromagnetic field enhancement in the nano-slit. The results could be attributed to the phenomenon of plasmon-exciton coupling with the effects of plasmon-induced energy transfer and light trapping.

PSI-plasmon interactions were previously demonstrated [110] to play a definite role in light propagation through a single microcavity slit patterned on a freestanding Au film. A monolayer of cysteine mutant of PSI was fabricated through covalent binding sulfhydryl moieties in the proteins to the gold grid. Multilayers were fabricated by crosslinking successive PSI layers. The study revealed a strong coupling between molecular excitations and surface plasmons, leading to macroscopic extended coherent states with increased temporal and spatial coherency, accompanied by a 13-fold enhancement of light transmission through the single-slit microcavity. It was proposed that the efficient long-range exciton transfer between the protein's pigments in the PSI antenna system could sustain efficient exciton transfer to the surface plasmon with minimal energy losses.

Furthermore, a potential optical switching system was demonstrated by applying a PSI- nanoparticle layer on microelectrodes. These PSI nanoparticles act as an antenna system, modulating and controlling charge distributions along microelectrodes [111]. Gold microelectrodes were coated by oriented multilayers of PSI and by PSI-gold and silver nanoparticles. Notably, the highest intensity, reaching four times that of uncoated electrodes, was observed in electrodes coated with PSI-silver nanoparticles. Furthermore, these intensities were further augmented upon the application of external voltage. It is suggested that the outstanding properties of light absorption and efficient exciton transfer of PSI antenna were enhanced by the plasmonic oscillations. This system holds promise for fast optical control in optical communications and Boolean logic gates.

Significant enhancements were observed in fluorescence intensity and photocurrent upon illumination of an oriented PSI assembly attached to a monolayer of graphene deposited on plasmonically active silver island film [37]. Nanoparticles were prepared through the reduction of silver nitrate in the presence of glucose. A monolayer of nanoparticles ranging in diameter from 70 to 140 nm and heights between 30 and 40 nm, was layered on flour tin oxide substrate covered with single-layer graphene functionalized with a moiety containing Co-nickel-nitrilotriacetic acid-pyrene and histag cytochrome $c$553, to which oriented layer of PSI from *Cyanidioschyzon merolae* was attached. At -300 mV, the current density reaches 224.4 nA cm$^{-2}$, representing more than a ~7-fold increase compared to the currents obtained without bias. The relatively low density of the PSI monolayer on the electrode likely contributed to the low current density. In the absence of external bias, the photocurrent output of around 5 nA cm$^{-2}$ was 22.6-fold higher than the reference. A fluorescence enhancement factor of ~9 was obtained with excitation at 550nm. The enhanced fluorescence was attributed to chlorophylls located at a distance larger than 8 nm from the nanoparticles. Previous observations [112] have indicated that the optimal distance for observation of metal-enhanced fluorescence is around 8 nm to 20 nm.

Enhancements of photocurrent were measured in single trimeric PSI complexes with photocurrent enhanced by the localized plasmons by a factor of 2.9 on average, reaching maximum enhancement values of up to 8. A strong plasmon enhancement was recorded in PSI located a few nm from the nano-rods. Hence, it was speculated that the strong dipole of the localized plasmon and its electric field could affect various steps of the electron transport process, including the charge separation and the individual electron-transfer steps along the acceptor chain.

Strong coupling was demonstrated as Rabi splitting in the transmission and fluorescence spectra due to light-matter interactions of the chlorophyll molecules in the photosystems [113]. In these experiments, cyanobacteria (*Synechococcus elongatus*) were placed in a microresonator. The Rabi splitting increased as a function of the increase in the chlorophyll *a* molecules, thus indicating the presence of a delocalized polaritonic state. Illumination caused splitting of the transmission band as a result of excitonic coupling between the pigments within photosystems and the cavity field through the formation of polaritonic state. It was suggested that this coupling could explain the very efficient photosynthetic energy transfer within the PSI of cyanobacteria.

**Plasmon-enhanced absorption and photocurrent in PSII electrochemical devices**
Enhancement of photocurrent by PSII immobilized on nanoparticles has also been demonstrated. Spinach PSII attached to 25 nm cystamine-2,6-dichlorobenzoquinone-modified Au nanoparticles generated a high photocurrent of 130 mA cm$^{-2}$ (35 mA cm$^{-2}$ mg$^{-1}$ chlorophyll), retaining the native oxygen evolution properties of PSII [114]. PSII bound to 20 nm gold nanoparticles through genetically engineered histag to the C-terminus of the CP47 subunit was earlier shown [115] to retain oxygen evolution. PSII monomers were attached to benzoquinone moiety through the $Q_B$ binding site of the protein and the PSII nanoparticles were placed on a graphite working electrode in a photochemical cell. Illumination at 530 nm, near the maximum absorption of the nanoparticles at 550 nm, yielded a 10-fold enhancement in maximum photocurrent in devices containing PSII-nanoparticles (1866 nA) relative to unconjugated PSII (203 nA), while conjugate Ag nanoparticles demonstrate only a 2x enhancement (430 nA). The enhanced photocurrent was a result of both improved light-harvesting in the 500-550 nm range not absorbed by chlorophyll *a* and direct electron injection facilitated through a nanoparticle ligand shell.

**Plasmon-enhanced fluorescence and electrochemical photocurrent in light-harvesting protein chlorophyll complexes devices**
An 18-fold enhancement of the chlorophyll fluorescence was observed for peridinin-chlorophyll-protein complexes when placed on a silver metal layer [116]. The monomer of the light-harvesting peridinin-chlorophyll-protein from dinoflagellate *Amphidinium carterae* was layered on silver island films on covered glass. Applying single-molecule spectroscopy it was found that "both the emission and absorption of photosynthetic antennae could be largely enhanced through plasmonic interactions. This enhancement was observed when exciting either chlorophyll or carotenoid and is attributed predominantly to an increase in the excitation rate in the antenna. The enhancement mechanism comes from the plasmon-induced amplification of electromagnetic fields inside the complex". Estimated enhancements of the fluorescence intensity of the Fenna-Matthews-Olson pigment-protein complex on silver island film are about 40-fold [37]. The protein from the photosynthetic green sulfur bacterium *Chlorobaculum tepidum* contains 7 (or 8) bacteriochlorophyll *a* molecules. The layering of the chlorophyll protein complex on silver island film enables the detection of fluorescence emission even when greatly diluted. The fluorescence was readily observed, even though the emission wavelength was displaced by over 300 nm from the maximum of the plasmon resonance of the silver island film layer.

Another work investigated the plasmonic effect of nanoparticles conjugated with light-harvesting complex II (LHCII) [117]. In these experiments "three types of core-shell metal TiO$_2$ nanoparticles with distinct surface plasmonic resonance were used. The plasmonic adsorption of the metal core enhanced the LHCII excitation through plasmon-induced resonance energy transfer. Charge separation was facilitated at the LHCII/TiO$_2$ interface, inducing quenching of the fluorescence and reduction of the fluorescence lifetime of LHCII". The plasmonic effects enhanced the photovoltaic properties after incorporation of the nanoparticles in LHCII-sensitized solar cells built on a 3D TiO$_2$ nano-tree photoanode. A TiO$_2$ nano-tree array on FTO glass was used as the photoanode. Iodine ions were used to shuttle electrons to the Pt cathode. The photocurrent was enhanced by two folds in the presence of the Ag and Au core-shell nanoparticles. The study reveals that the core-shell nanoparticles can enhance the LHCII-sensitized solar cells based on multiple mechanisms, including enhancing light harvesting, promoting plasmon-induced resonance energy transfer from nanoparticles to LHCII, and facilitating charge injection from excited LHCII into $TiO_2$.

**Plasmon-enhanced fluorescence in PSI**
Metal nanoparticle-enhanced fluorescence of molecules can be attributed to either by an electric field effect or a local plasmon effect [118]. The electric field amplifies the emission intensity of the molecules by accelerating the excitation rate. Nanoparticles augment the local electric field, leading to higher fluorescence intensity without altering the fluorescence lifetime [119]. Maximum enhancements observed are 36 for PSI-silver island film and 37 for PSI-Au nanoparticles, with average enhancements of 7 and 9, respectively [120]. In single-molecule experiments, PSI from *Thermosynechococcus elongatus* was added to gold nanoparticles or placed onto silver island film. Both nanostructures elicited an increase in fluorescence from the higher energy chlorophylls, consequently reducing exciton transfer to the (P700) reaction center.

Plasmon-enhanced single-particle fluorescence spectroscopy was used to decipher the origin of both short- and long-wavelength emitting species in monomeric PSI from *Thermosynechococcus elongatus* at room temperature [121]. Monomeric PSI were immobilized on the nanorodes and coated by poly(vinyl alcohol). Plasmonic enhancement boosted fluorescence by a factor of 13-14. Applying quantum chemical calculations, together with reported X-ray structural and spectroscopic data, implicated chlorophyll a monomer A3 as a candidate for the short-wavelength emitter and the B31-B32 chlorophyll dimer as a candidate for the long-wavelength emitter.

Plasmon-induced enhancement of fluorescence has also been investigated in *Chlorobaculum tepidum* reaction centers [122]. Steady-state fluorescence measurements of the reaction centers deposited on silver island film substrates revealed fluorescence intensity enhancements of up to 60 times, with strong excitation wavelength dependence. Time-resolved experiments showed significantly shortened fluorescence lifetimes of reaction centers, indicating that plasmons enhance not only the excitation and emission but also the energy transfer pathways among the pigments of the reaction center.

The effect of raft aggregates of silver metal nanoparticles on the enhancement of fluorescence was investigated in thin film PSI trimer complexes from the cyanobacteria *Synechococcus leopoliensis* [123]. Layers of PSI were deposited on a glass-coated layer of 20 nm diameter Ag nanoparticles. With far-field illumination a 5- to 20-fold fluorescence enhancement was observed for PSI complexes adjacent to nanoparticles, with the most prominent enhancement detected near aggregated nanoparticles. These findings align with the earlier observation of an increase in the absorption enhancement in PSI-nanoparticle aggregates [105].

Plasmon-enhanced fluorescence of PSI complex with their peripheral light-harvesting antenna (LHCI) from *Cyanidioschyzon merolae* and of PSI complex from *Thermosynechococcus elongatus* was measured [37]. The PSI complexes were spin-coated on a silver island substrate. The plasmon enhancement factor ranged between 1 and 5 depending on the type of silver island film preparation for both complexes. It was concluded that the plasmon enhancement factor heavily depends on the type of interaction between the nanoparticles and the chromophores.

**Plasmon-enhanced absorption and photocurrent in electrochemical RC devices**
In the study of Friebe *et al*. [87], bacterial reaction center light-harvesting 1(RC-LH1) complexes and cytochrome c were deposited on a nanostructured silver substrate. Electrochemical measurement, conducted with 2,3-dimethoxy-5-methyl-p-benzoquinone as an electron transport mediator, revealed significant enhancement in photocurrent. Under 1 sun illumination, the cell yielded a peak photocurrent of 166 $\mu A\ cm^{-2}$ while under 4 suns, it reached a record maximum of 416 $\mu A\ cm^{-2}$. The observed enhancements can be attributed to "a 2.5-fold plasmonic enhancement of light absorption per RC-LH1 complex on the rough silver substrate". This plasmon-enhanced fluorescence was due to the increase in fluorescence yield, and radiative rate of the RC-LH1 complexes. Furthermore, nanostructuring of the silver substrate also contributed to the stability of the protein under continuous illumination, by almost an order of magnitude relative to a nanostructured bulk silver.

**Spatial modulation of light transmission through a microcavity by PSI**
The work of Carmeli *et al*. [110]. elucidated the intricate coupling between light transmission and PSI complexes within microcavity patterns. This was achieved by covalently bonding oriented multilayers of PSI to a thin Au film pattern featuring microcavities. The transmitted light exhibited a novel wave pattern, accompanied by a remarkable 13-fold increase in intensity. This study lays the foundation for the development of bio-inspired photonic devices.

**CONCLUSIONS**
This article discusses advancement in the fabrication of electronic junctions in photosynthetic protein-solid state devices, along with the theory of current flow through these proteins and junctions, as well as the theory of light-matter interactions. It also discusses the effect of light-matter interactions on energy levels and current flow within photosynthetic protein-nano-structure hybrids. Although research into the light-matter interaction of nano-structure-photosynthetic protein hybrids is still in its early stages the novel findings hold significant promise. Remarkable phenomena emerge when the photosynthetic proteins are exposed to plasmon resonance. These chlorophyll-protein complexes exhibit unique capabilities, such as the coherent transfer of excitons by the antenna of chlorophyll molecules and the conversion of absorbed photons to electric voltage with ~100% quantum efficiency by the reaction centers. The plasmon resonance enhances absorption, fluorescence, electron transport rates, modulation of light transmission through nano-slits, and charge modulation on microelectrodes. However, current studies mostly show weak coupling between plasmons and the protein. Initial studies suggest that nanostructures can be fine- tuned to match the energy levels of the photosynthetic proteins, facilitating strong coupling. Practical applications of Rabi splitting of the absorption maxima offer potential avenues for engineering diverse bio-photonic devices based on various photosynthetic proteins.

Considerable progress has been made in the fabrication and efficiency of electronic junctions in solid-state and electrochemical devices. Several of the devices reached high internal quantum efficiency. However, the external quantum efficiency remains low due to either low absorption cross section or inadequate electronic junctions in the hybrid devices. The rather short lifetime of photosynthetic proteins particularly PSII, in electrochemical devices primarily stems from the corrosive effects of oxygen radicals generated during light excitation of the pigments and electron transport in the photosynthetic reaction centers. The utilization of redox polymers and hydrogels improves the device's lifespan and offers ways for future improvement. In contrast to solution-phase applications, photosynthetic proteins in solid-state hybrid devices can function for extended durations, owing to the scarcity of oxygen radicals and the absence of chemical degradation in the dry environment. The major drawback of solid-state devices lies in their low absorption cross-section; however, the fabrication of oriented multilayers of the protein complexes presents a means to enhance light absorption in these devices.

**CONFLICT OF INTEREST STATEMENT**
The authors declare no competing financial interests.

**REFERENCES**

1. Ebbesen TW., 2016, Acc. Chem. Res., 49, 2403-12
2. Cao E, Lin W, Sun M, Liang W, Song Y., 2018, Nanophot., 7, 145-67
3. Hutchison JA, Liscio A, Schwartz T, Canaguier-Durand A, Genet C, Palermo, V, Samorì, P, Ebbesen, TW, 2013, Adv. Mater., 25, 2481-5
4. Bostick CD, Mukhopadhyay S, Pecht I, Sheves M, Cahen D, Lederman D., 2018, Rep. Prog. Phys., 81, 026601
5. Katz E, Willner I., 2004, Ang. Chem. Inter. Ed., 43, 6042-108
6. Vigneshvar S, Sudhakumari CC, Senthilkumaran B, Prakash H. 2016. Fron. Bioengin. Biotechnol., 4
7. Willner I, Yan YM, Willner B, Tel-Vered R. 2009, Fuel Cell., 9, 7-24
8. Badura A, Guschin D, Kothe T, Kopczak MJ, Schuhmann W, Rögner M. 2011, Ener. Environ. Sci., 4, 2435-40
9. Nguyen K, Bruce BD. 2014, Biochim. Biophy. Acta-Bioener., 1837, 1553-66
10. Yehezkeli O, Tel-Vered R, Michaeli D, Willner I, Nechushtai R. 2014, Photosyn. Res., 120, 71-85
11. Milano F, Punzi A, Ragni R, Trotta M, Farinola GM. 2019, Adv. Func. Mater., 29, 1805521
12. Zhang JZ, Reisner E. 2020, Nat., Rev. Chem., 4: 6-21
13. Proppe AH, Li YGC, Aspuru-Guzik A, Berlinguette CP, Chang CJ, Cogdell, R, Doyle, AG, Flick, J, Gabor, NM, van Grondelle, R, Hammes-Schiffer, S, Jaffer, SA, Kelley, SO, Leclerc, M, Leo, K, Mallouk, TE, Narang, P, Schlau-Cohen, GS, Scholes, GD, Vojvodic, A, Yam, VWW, Yang, JY, Sargent, EH, 2020, Nat, Rev. Mater., 5: 828-46
14. Tel-Vered R, Willner I., 2014. Chemelectrochem. 1, 1778-97
15. Teodor AHB, B. D., 2020, T. Btech., 38, 1329-42
16. Wolfe KD, Dervishogullari D, Stachurski CD, Passantino JM, Jennings GK, Cliffel DE., 2020,

Chemelectrochem., 7
17. Weliwatte NS, Grattieri M, Minteer SD., 2021, Photochem. Photobiol. Sci., 20, 1333-56
18. Kim YJ, Hong H, Yun J, Kim SI, Jung HY, Ryu W., 2021, Adv. Mater., 33, 2005919
19. Blankenship RE, Tiede DM, Barber J, Brudvig GW, Fleming G, Ghirardi, M, Gunner, MR, Junge, W, Kramer, DM, Melis, A, Moore, TA, Moser, CC, Nocera, DG, Nozik, AJ, Ort, DR, Parson, WW, Prince, RC, Sayre, RT, 2011, Science, 332, 805-9
20. Wang C, O'Hagan MP, Willner B, Willner I., 2022, Chem.-a Eur. J., 28,
21. Wolfe KD, Dervishogullari D, Passantino JM, Stachurski CD, Jennings GK, Cliffel DE., 2020, Curr. Opin. Electrochem., 19, 27-34
22. Ozboyaci M, Kokh DB, Corni S, Wade RC., 2016. Quart. Rev. Biophy., 49, e4
23. Moser CC, Dutton PL., 1992, Biochim, Biophy. Acta, 1101. 171-6
24. Winkler JR, Gray HB., 2014, Chem. Rev., 114, 3369-80
25. Amdursky N, Marchak D, Sepunaru L, Pecht I, Sheves M, Cahen D., 2014, Adv.Mater., 26, 7142-61
26. Panda SS, Katz HE, Tovar JD, 2018, Chem. Soc. Rev., 47, 3640-58
27. Marcus RA, Sutin N, 1985, Biochim. Biophys. Acta, 811, 265-322
28. Landauer R., 1957, IBM J. Res. Develop., 1, 223-31
29. Frolov L, Rosenwaks Y, Carmeli C, Carmeli I., 2005, Adv.Mater., 17, 2434-7
30. Heifler O, Carmeli C, Carmeli I., 2018, J. Phys.l Chem. C, 122, 11550-6
31. Carmeli I, Mangold M, Frolov L, Zebli B, Carmeli C, Richter, S, Holleitner, AW, 2007, Adv. Mater., 19: 3901-9308
32. Yehezkeli O, Tel-Vered R, Wasserman J, Trifonov A, Michaeli D, Nechushtai, R, Willner, I, 2012, Nat. Comm., 3
33. Mershin A, Matsumoto K, Kaiser L, Yu DY, Vaughn M, Nazeeruddin, MK, Bruce, BD, Graetzel, M, Zhang, SG, 2012. Sci. Report, 2
34. Lebedev N, Trammell SA, Tsoi S, Spano A, Kim JH, Xu, J, Twigg, ME, Schnur, JM, 2008, Lang., 24, 8871-6
35. Kothe T, Pöller S, Zhao F, Fortgang P, Rögner M, Schuhmann, WN, Plumeré, N, 2014, Chem.– A Eur. J., 20, 11029-34
36. Proux-Delrouyre V, Demaille C, Leibl W, Sétif P, Bottin H, Bourdillon C., 2003, J. Am.Chem. Soc., 125, 13686-92
37. Kowalska D, Szalkowski M, Sulowska K, Buczynska D, Niedziolka-Jonsson J, Kargul, J, Lokstein, H, Mackowski, S, 2020, Int.
J. Mol. Sci., 21
38. Friebe VM, Delgado JD, Swainsbury DJK, Gruber JM, Chanaewa A, van Grondelle, R, von Hauff, E, Millo, D, Jones, MR, Frese, RN, 2016, Adv. Funct. Mater., 26, 285-92
39. Malavath T, Caspy I, Netzer-El SY, Klaiman D, Nelson N., 2018, Biochim. Biophy. Acta-Bioener., 1859, 645-54
40. Loll B, Kern J, Saenger W, Zouni A, Biesiadka J. 2005, Nature, 438, 1040-4
41. Stowell MH, McPhillips TM, Rees DC, Soltis SM, Abresch E, Feher G., 1997, Science, 276, 812-6
42. Brettel K, Leibl W., 2001, Biochim. Biophys. Acta-Bioener., 1507, 100-14

43. Blankenship, R. E., The Basic Principles of Photosynthetic Energy Storage, 2008 in Molecular Mechanisms of Photosynthesis, Blankenship, R. E. (Ed) Wily, on line, 1-10
44. Nelson N, Junge W., 2015, Annu. Rev. Biochem., 84, 659-83
45. Akhtar P, Caspy I, Nowakowski PJ, Malavath T, Nelson N, Tan H.-S., Lambrev, PH. 2021, J. Am. Chem. Soc.,143, 14601-12
46. Toporik H, Carmeli I, Volotsenko I, Molotskii M, Rosenwaks Y, Carmeli, C, Nelson, N., 2012, Adv. Mater., 24, 2988-91
47. Carmeli I, Leitus G, Naaman R, Reich S, Vager Z., 2003, Israel J. Chem. 43, 399-405
48. Heifler O, Carmeli C, Carmeli I. 2020, Lang., 36, 4556-62
49. Sepunaru L, Tsimberov I, Forolov L, Carmeli C, Carmeli I, Rosenwaks Y., 2009, Nano Lett., 9. 2751-5
50. Jordan P, Fromme P, Witt HT, Klukas O, Saenger W, Krauss N., 2001. Nature 411: 909-17
51. Gerster D, Reichert J, Bi H, Barth JV, Kaniber SM, Holleitner, AW, Visoly-Fisher, I, Sergani, S, Carmeli, I, 2012, Nat. Nanotech., 7, 673-6
52. Barhom H, Carmeli C, Carmeli I., 2021, J. Phys. Chem. B, 125, 722-8
53. Barhom H., 2016, Photosystem I from Cyanobacteria As a Basis for The Fabrication of Solid State Photo-voltaic Devices. Tel Aviv University, Tel Aviv. 86 pp.

54. Mukherjee D, May M, Khomami B., 2011, J. Coll. Interf. Sci., 358, 477-84
55. Castañeda Ocampo OE, Gordiichuk P, Catarci S, Gautier DA, Herrmann A, Chiechi RC., 2015, J. Am. Chem. Soc., 137, 8419-27
56. Sepunaru L, Pecht I, Sheves M, Cahen D., 2011, J. Am. Chem. Soc. 133, 2421-3
57. Carmeli I, Kumar KS, Heifler O, Carmeli C, Naaman R., 2014, Ang, Chem. Inter. Ed., 53, 8953-8
58. Naaman R, Paltiel Y, Waldeck DH., 2019, Nat. Rev. Chem., 3, 250-60
59. Büttiker M., 1988, Phys. Rev. B, 38, 9375-89
60. Gordiichuk PI, Wetzelaer GJAH, Rimmerman D, Gruszka A, de Vries JW, Saller, M, Gautier, D A, Catarci, S, Pesce, D, Richter, S, Blom, P W M, Herrmann, A, 2014, Adv. Mater., 26, 4863-4869
61. Yao K, Jiao HF, Xu YX, He QQ, Li F, Wang XF., 2016, J. Mater. Chem. A, 4, 13400-6
62. Frolov L, Wilner O, Carmeli C, Carmeli I., 2008, Adv. Mater. 20, 263-269
63. Das R, Kiley PJ, Segal M, Norville J, Yu AA, Wang, LY, Trammell, SA, Reddick, LE, Kumar, R, Stellacci, F, Lebedev, N, Schnur, J, Bruce, BD, Zhang, SG, Baldo, M, 2004, Nano Lett, 4, 1079-83
64. Lee JW, Lee I, Greenbaum E., 1996, Biosens. Bioelect. 11, 375-87
65. Efrati A, Lu CH, Michaeli D, Nechushtai R, Alsaoub S, Schuhmann, W, Willner, I, 2016, Nat. Ener., 1
66. Miyachi M, Yamanoi Y, Shibata Y, Matsumoto H, Nakazato K, Konno, M, Ito, K, Inoue, Y, Nishihara, H, 2010, Chem. Comm. 46, 2557-9
67. Ciobanu M, Kincaid HA, Lo V, Dukes AD, Jennings GK, Cliffel DE., 2007, J. Electroanal. Chem. 599, 72-8
68. Ciesielski PN, Hijazi FM, Scott AM, Faulkner CJ, Beard L, Emmett, K, Rosenthal, SJ, Cliffel, D, Jennings, GK, 2010, Biores. Technol. 101, 3047-53
69. Wolfe KD, Gargye A, Mwambutsa F, Then L, Cliffel DE, Jennings GK., 2021, Lang. 37, 10481-9
70. Gunther D, LeBlanc G, Prasai D, Zhang JR, Cliffel DE, Bolotin, KI, Jennings, GK, 2013, Lang. 29, 4177-80
71. Feifel SC, Stieger KR, Lokstein H, Lux H, Lisdat F., 2015, J. Mater. Chem. A 3, 12188-96
72. Morlock S, Subramanian SK, Zouni A, Lisdat F., 2021, ACS Appl. Mater. Interf. 13, 11237-46
73. Zhao FY, Wang PP, Ruff A, Hartmann V, Zacarias S, Pereira, IAC, Nowaczyk, MM, Rogner, M, Conzuelo, F, Schuhmann, W, 2019, Ener. Environ. Sci. 12, 3133- 43
74. Wang P, Zhao F, Frank A, Zerria S, Lielpetere A, Ruff, A, Nowaczyk, MM, Schuhmann, W, Conzuelo, F, 2021, Adv. Ener. Mater. 11: 2102858
75. Kato M, Cardona T, Rutherford AW, Reisner E., 2013, J. Am. Chem. Soc. 135, 10610-3
76. Fang X, Sokol KP, Heidary N, Kandiel TA, Zhang JZ, Reisner E., 2019, Nano Lett. 19, 1844-50
77. Mersch D, Lee CY, Zhang JZ, Brinkert K, Fontecilla-Camps JC, Rutherford, AW, Reisner, E, 2015, J Am Chem Soc 137, 8541-9
78. Sokol KP, Mersch D, Hartmann V, Zhang JZ, Nowaczyk MM, Rogner, M, Ruff, A, Schuhmann, W, Plumere, N, Reisner, E, 2016, Ener. Environ. Sci. 9, 3698-709
79. Badura A, Esper B, Ataka K, Grunwald C, Wöll C, Kuhlmann, J, Heberle, J, Rögner, M, 2006, Photochem Photobiol 82, 1385- 90
80. Voloshin, R.A., Shumilova, S.M., Zadneprovskaya, E.V., Zharmukhamedov, S.K., Alwasel, V, S., Hou, H.J.M., Allakhverdiev, S.I., Photosystem II in Bio-photovoltaic Devices, Photosyn. 60(1) (2022) 121-135
81. Yehezkeli O, Tel-Vered R, Michaeli D, Willner I, Nechushtai R., 2013, Photosynth. Res 116, 1-20
82. Nakamura C, Hasegawa M, Yasuda Y, Miyake J., 2000, Appl. Biochem. Biotechnol. 84-6, 401-8
83. Trammell SA, Wang L, Zullo JM, Shashidhar R, Lebedev N., 2004, Biosen.s& bioelect. 19, 1649-55
84. Lebedev N, Trammell SA, Spano A, Lukashev E, Griva I, Schnur J, 2006, J. Am. Chem. Soc. 128, 12044-5
85. Paddock ML, Weber KH, Chang C, Okamura MY, 2005, Biochem. 44, 9619-25
86. Gebert J, Reiner-Rozman C, Steininger C, Nedelkovski V, Nowak C, Wraight, CA,.Naumann, RLC, 2015, J. Phys. Chem. C 119, 890-5
87. Friebe VM, Millo D, Swainsbury DJK, Jones MR, Frese RN., 2017, ACS Appl. Mater. Interf. 9, 23379-88

88. Friebe VM, Barszcz AJ, Jones MR, Frese RN., 2022, Ang. Chem.-Inter. Ed. 61
89. Vitukhnovskaya LA, Zaspa AA, Semenov AY, Mamedov MD., 2023, Biochim. Biophys. Acta - Bioener. 1864, 148975
90. Terasaki N, Yamamoto N, Hattori M, Tanigaki N, Hiraga T, Ito, K, Konno, M, Iwai, M, Inoue, S. Uno, Y, Nakazato, K, 2009. Lang. 25, 11969-74
91. Terasaki N, Yamamoto N, Tamada K, Hattori M, Hiraga T, Tohri, A, Sato, I, Iwai, M, Iwai, M, Taguchi, S, Enami, I, Inoue, Y, Yamanoi, Y, Yonezawa, T, Mizuno, K, Murata, M, Nishihara, H, Yoneyama, S, Minakata, M, Ohmori, T, Sakai, M, Fujii, M, 2007, Biochim. Biophys. Acta-Bioener. 1767, 653-9
92. Takshi A, Yaghoubi H, Wang J, Jun D, Beatty JT., 2017, Biosens. (Basel) 7
93. Chikkaraddy R, de Nijs B, Benz F, Barrow SJ, Scherman OA, Rosta, E, Demetriadou, A, Fox, P, Hess, O, Baumberg, JJ, 2016, Nature 535, 127-30
94. Zeng P, Cadusch J, Chakraborty D, Smith TA, Roberts A, Sader, JE, Davis, TJ, Gómez, DE, 2016, Nano Lett.16, 2651-6
95. Campion A, Kambhampati P., 1998, Chem. Soc. Rev. 27, 241-50
96. Schaadt DM, Feng B, Yu ET., 2005, Appl. Phys. Lett. 86, 063106
97. Okamoto K, Funato M, Kawakami Y, Tamada K, 2017, J. Photochem. Photobiol. C, Photoch. Rev, 32, 58-77
98. Kneipp K, Wang Y, Kneipp H, Perelman LT, Itzkan I, Dasari, RR, Feld, MS, 1997, Phys. Rev. Lett. 78, 1667- 70
99. Cotton TM, Schultz SG, Van Duyne RP., 1980, J. Am. Chem. Soc. 102: 7960-2
100. Osawa M, Matsuda N, Yoshii K, Uchida I., 1994, J. Phys. Chem. 98, 12702-7
101. Fu Y, Lakowicz JR., 2009, Laser Photon Rev 3, 221-32
102. Spinelli P, Polman A., 2012, Optic. Expr. 20, A641-A54
103. Atwater HA, Polman A., 2010, Nat. Mater. 9, 205-13
104. Zhong X, Chervy T, Wang S, George J, Thomas A, Hutchison, JA, Devaux, E, Genet, C, Ebbesen, TW, 2016, Ang. Chem. Internat. Ed. 55, 6202-6
105. Carmeli I, Lieberman I, Kraversky L, Fan ZY, Govorov AO, Markovich, G, Richter, S, 2010, Nano Lett. 10, 2069- 74
106. Govorov AO, Carmeli I., 2007, Nano Lett. 7, 620-5
107. Niroomand H, Pamu R, Mukherjee D, Khomami B., 2018, Mrs Comm. 8, 823-9
108. Pamu R, Lawrie BJ, Khomami B, Mukherjee D., 2021, ACS Appl. Nano Mater. 4, 1209-19
109. Zheng Zeng TM, Wendi Zhang, Bhawna Bagra, Zuowei Ji, Ziyu Yin, Kokougan Allado, and Jianjun Wei, 2018, Appl. Bio Mater. 1, 802–7
110. Carmeli I, Cohen M, Heifler O, Lilach Y, Zalevsky Mujica, ZV, Richter, S, 2015, Nat. Comm. 6
111. Carmeli I, Tanriover I, Malavath T, Carmeli C, Cohen Abulafia, MY, Girshevitz, O, Richter, S, Aydin, K, Zalevsky, Z, 2023, ACS Appl. Nano Mater. 6, 13668-76
112. Pascal AA, Liu Z, Broess K, van Oort B, van Amerongen H, Wang, C, Horton, P, Robert, B, Chang, W, Ruban, A, 2005, Nature 436, 134-7
113. Rammler T, Wackenhut F, zur Oven-Krockhaus S, Rapp J, Forchhammer K, Harter, K, Meixner, AJ, 2022, J. Biophot. 15, e202100136
114. Shoyhet H, Pavlopoulos NG, Amirav L, Adir N., 2021, J, Mater, Chem, A 9, 17231-41
115. Noji T, Suzuki H, Gotoh T, Iwai M, Ikeuchi M, Tomo, T, Noguchi, T, 2011, J. Phys. Chem. Lett. 2, 2448-52
116. Mackowski S, Wörmke S, Maier AJ, Brotosudarmo TH, Harutyunyan H, Hartschuh, A, Govorov, AO, Scheer, H, Bräuchle, C, 2008, Nano Lett 8, 558-64
117. Yang Y, Gobeze HB, D'Souza F, Jankowiak R, Li J., 2016, Adv. Mater. Interf. 3, 1600371
118. Lakowicz JR, Chowdhury MH, Ray K, Zhang J, Fu Y, Badugu, R, Sabanayagam, CR, Nowaczyk, K, Szmacinski, H, Aslan, K, Geddes, CD, 2006, Proc SPIE Int Soc Opt Eng 6099, 609909
119. Ashraf I, Konrad A, Lokstein H, Skandary S, Metzger M, Djouda, JM, Maurer, T, Adam, PM, Meixner, AJ, Brecht, M, 2017, Nanoscale 9, 4196-204
120. Nieder JB, Bittl R, Brecht M., 2010, Ang. Chem.-Inter. Ed. 49, 10217-20
121. Hatazaki S, Sharma DK, Hirata S, Nose K, Iyoda T, Kölsch, A, Lokstein, H, Vacha, M, 2018, J Phys Chem Lett 9, 6669-75
122. Maćkowski S, Czechowski N, Ashraf KU, Szalkowski M, Lokstein H, Cogdell, RJ, Kowalska, D. 2016, FEBS Let. 590, 2558-65
123. Kim I, Bender SL, Hranisavljevic J, Utschig LM, Huang L, Wiederrecht, GP. Tiede, DM, 2011. Nano Lett. 11, 3091-8